%% file: main.tex
\pdfoutput=1

\def\paperversion{1} % 0 - anonymous, 1 - camera ready, 2 - full version
\def\style{0} % 0 - article, 1 - lncs, 2 - acm, 3 - usenix, 4 - ieeetr, 5 - sciendo, 6 - lipics
\def\showcomments{1} % 0 - no comments, 1 - show comments

\ifnum\showcomments=0
\PassOptionsToPackage{disable}{todonotes}
\fi
\def\anonymous{0}
\def\cameraready{1}
\def\fullversion{2}

\def\article{0}
\def\lncs{1}
\def\sigalternate{2}
\def\usenix{3}
\def\ieeetr{4}
\def\sciendo{5}
\def\lipics{6}

\PassOptionsToPackage{protrusion=true, expansion, tracking, kerning, spacing}{microtype}
\PassOptionsToPackage{usenames,dvipsnames,table}{xcolor}

\ifnum\style=\article
	\documentclass[10pt,A4,headings=standardclasses,numbers=noenddot]{scrartcl}
	\usepackage{amsmath,amsfonts,amssymb,amsthm}
	\usepackage{fullpage}
    \usepackage[backend=bibtex,style=alphabetic]{biblatex}
	\newtheorem{theorem}{Theorem}[section]
	\let\oldproof\proof
	\let\oldendproof\endproof
	\renewenvironment{proof}{
	\oldproof
	}{%\pushQED{\qed}
	\oldendproof}
	\newcommand\keywords{}

\fi

\ifnum\style=\lncs
	\let\accentvec\vec % countermeasure against "unable to redefine math accent \vec
	\documentclass[10pt]{styles/llncs}
	\usepackage{amssymb}
	\usepackage{expl3}
	\let\vec\accentvec % countermeasure
	\newcommand\addbibresource[1]{
	 	\newcommand\bibfiles{#1}
	}
	\newcommand\printbibliography{
		\ifnum\paperversion=\fullversion
					\bibliographystyle{alpha}
		\else
					\bibliographystyle{splncs04}
		\fi
		\bibliography{\bibfiles}
	}
\fi

\ifnum\style=\sigalternate
    \ifnum\paperversion=\anonymous
        \documentclass[sigconf,anonymous,natbib=false]{acmart}
        \setcopyright{none}
        \renewcommand\footnotetextcopyrightpermission[1]{}
    \fi
    \ifnum\paperversion=\cameraready
        \documentclass[sigconf,natbib=false]{acmart}
        \setcopyright{acmlicensed}
        \copyrightyear{2019}
        \acmYear{2019}
        \acmConference[CCS '19]{2019 ACM SIGSAC Conference on Computer and Communications Security}{November 11--15, 2019}{London, United Kingdom}
        \acmBooktitle{2019 ACM SIGSAC Conference on Computer and Communications Security (CCS '19), November 11--15, 2019, London, United Kingdom}
        \acmPrice{15.00}
        \acmDOI{xxx}
        \acmISBN{xxx}
    \fi
    \ifnum\paperversion=\fullversion
        \documentclass[sigconf,natbib=false]{acmart}
        \setcopyright{none}
        \renewcommand\footnotetextcopyrightpermission[1]{}
    \fi
	\newtheorem{remark}{Remark}
    \usepackage[
    backend=bibtex,
    style=ACM-Reference-Format
    ]{biblatex}
\fi

\ifnum\style=\usenix
	\documentclass[letterpaper,twocolumn,10pt]{article}
	\usepackage{styles/usenix}
	\usepackage{epsfig}
	\usepackage{amsmath,amsfonts, amssymb, amsthm}
	\newtheorem{theorem}{Theorem}[section]
	\newtheorem{definition}[theorem]{Definition}
	\newtheorem{remark}{Remark}
	\newtheorem{lemma}{Lemma}
	\let\oldproof\proof
	\let\oldendproof\endproof
	\renewenvironment{proof}{
	\oldproof
	}{\pushQED{\qed}\oldendproof}
	\usepackage[
    backend=bibtex,
    style=numeric
    ]{biblatex} % Need to comment out \usepackage{cite} in usenix.sty
	\DeclareMathAlphabet{\mathcal}{OMS}{cmsy}{m}{n}
\fi

\ifnum\style=\ieeetr
	\documentclass[compsoc, conference, letterpaper, 10pt, times]{styles/IEEEtran}
	\IEEEoverridecommandlockouts
    \def\BibTeX{{\rm B\kern-.05em{\sc i\kern-.025em b}\kern-.08em
            T\kern-.1667em\lower.7ex\hbox{E}\kern-.125emX}}
		\newtheorem{theorem}{Theorem}[section]
    \newtheorem{definition}[theorem]{Definition}
    \newtheorem{remark}{Remark}
    \newtheorem{proof}{Proof}
    \newtheorem{lemma}{Lemma}
    \usepackage[style=ieee,backend=bibtex]{biblatex}
\fi

\ifnum\style=\sciendo
\documentclass[USenglish,oneside,twocolumn]{article}
\PassOptionsToPackage{english}{babel}
\usepackage[big]{styles/dgruyter_NEW} %Contains amsmath, amssymb, authblk, multicol, graphicx, babel,  inter alia
\usepackage{amsfonts,amsthm}
\usepackage[
style=numeric-comp,
backend=bibtex8,
maxbibnames=99,
natbib=true,
maxcitenames=6,
giveninits=true,
]{biblatex}
\cclogo{\includegraphics{styles/by-nc-nd.pdf}}

\newtheorem{theorem}{Theorem}[section]
\allowdisplaybreaks
\fi

\ifnum\style=\lipics
\ifnum\paperversion=\anonymous
\documentclass[a4paper,UKenglish,numberwithinsect,thm-restate,anonymous]{styles/lipics-v2019}
\else
\documentclass[a4paper,UKenglish,numberwithinsect,thm-restate]{styles/lipics-v2019}
\fi
\PassOptionsToPackage{english}{babel}

\newcommand\printbibliography{
  \bibliographystyle{plainurl}% the mandatory bibstyle
  \bibliography{\bibfiles}
}

\newcommand\addbibresource[1]{
  \newcommand\bibfiles{#1}
}

\fi

\input{0-packages}

\newaliascnt{conj}{theorem}
\crefname{conjecture}{Conjecture}{Conjectures}

\newtheorem{definition}[theorem]{Definition}
\newtheorem{lemma}[theorem]{Lemma}

\newtheorem{conjecture}[theorem]{Conjecture}
\newtheorem{corollary}[theorem]{Corollary}
\newtheorem{remark}[theorem]{Remark}

\theoremstyle{proof}

\ifnum\style=0
  \usepackage{authblk}

\fi
\input{0-macros}

\ifnum\style=5

\journalname{Proceedings on Privacy Enhancing Technologies}
\DOI{Editor to enter DOI}
\startpage{1}
\received{..}
\revised{..}
\accepted{..}

\journalyear{..}
\journalvolume{..}
\journalissue{..}

\fi

\ifnum\style=2
    \ifnum\paperversion=0
        \acmConference[Anonymous Submission to ACM CCS ???]{ACM Conference on Computer and Communications Security}{Due [Date]}{[City], [Country]}
        \acmYear{[Year]}
    \fi
    \ifnum\paperversion=1
      \setcopyright{acmlicensed}
      \copyrightyear{2019}
      \acmYear{2019}
      \acmConference[CCS '19]{2019 ACM SIGSAC Conference on Computer and Communications Security}{November 11--15, 2019}{London, United Kingdom}
      \acmBooktitle{2019 ACM SIGSAC Conference on Computer and Communications Security (CCS '19), November 11--15, 2019, London, United Kingdom}
      \acmPrice{15.00}
      \acmDOI{10.1145/3319535.3354262}
      \acmISBN{978-1-4503-6747-9/19/11}
    \fi
    \ifnum\paperversion=2
      \renewcommand\footnotetextcopyrightpermission[1]{}
      \let\oldmakealltitles\makealltitles
      \renewcommand\makealltitles{
        \oldmakealltitles
        \thispagestyle{fancy}
        \lhead{An extended abstract of this paper appears at ACM CCS '19. This is the full version.}
        \rhead{}
        \pagestyle{plain}
      }
    \fi
\fi

\date{April 11, 2022}

\ifnum\style=2
  \title[Short Title]{Title}
  \subtitle{Subtitle}
\else
    \title{On Defeating Graph Analysis of \\ Anonymous Transactions
    }
  \ifnum\style=5
  \fi
\fi

\ifnum\paperversion=2
  \ifnum\style=0
    \titlehead{An extended abstract of this paper appears at ACM CCS '19. This is the full version.}
  \else
    \ifnum\style<2
    \subtitle{\textnormal{(Full Version)}}
    \fi
  \fi
\fi

\ifnum\paperversion=0
    \author{Anonymous Author(s)}
\else
    \author[1]{Christoph Egger}
    \author[1]{Russell W. F. Lai}
    \author[1]{Viktoria Ronge}
    \author[2]{Ivy K. Y. Woo}
    \author[3]{Hoover H. F. Yin}
    \affil[1]{Friedrich-Alexander-Universit{\"a}t Erlangen-N{\"u}rnberg}
    \affil[2]{Independent}
    \affil[3]{The Chinese University of Hong Kong}
\fi

\newuser{V}{Vicky}{green}
\newuser{S}{Who}{yellow}
\newuser{I}{Ivy}{RoyalBlue}
\newuser{R}{Russell}{yellow}
\newuser{C}{Christoph}{RoyalPurple}

\ifnum\style=1
\else
\ifnum\style=6
\else
\fi
\fi

\begin{document}
\ifnum\showcomments=1
\fi

\ifnum\style=2
\ifnum\paperversion=1
% If style=2 - acm generate CCS Code (ACM Computing Classification Scheme)
% from http://dl.acm.org/ccs.cfm  and paste it here.
% Caution: No indent before \begin{CCSXML}


\ccsdesc[500]{Security and privacy~Cryptography}
\fi
\fi

% Use this commands in that order independet of the style. It will fix
% everything.
\allabstracts{\input{0-abstract}}
\allkeywords{anonymous cryptocurrencies, ring signatures, random directed graph connectivity}
\makealltitles
\ifnum\paperversion=2
% \setcounter{tocdepth}{1}
% \tableofcontents
\fi
\ifnum\showcomments=1
% \tableofcontents
\fi

\hypersetup{
    pdftitle = {On Defeating Graph Analysis of Anonymous Transactions},
    pdfauthor = {Christoph Egger, Russell W. F. Lai, Viktoria Ronge, Ivy K. Y. Woo, Hoover H. F. Yin}
  }

% --- -----------------------------------------------------------------
% --- Input Files by Section.
% --- -----------------------------------------------------------------

% \inputanon{FILENAME} include FILENAME only in paperstyle=anonymous
% \inputcam{FILENAME} include FILENAME only in paperstyle=cameraready
% \inputfull{FILENAME} include FILENAME only in paperstyle=fullversion

\input{intro}

\input{prelim}

%\input{new_part}

\input{connectivity}

\input{partitioning}

\input{conjectures}

\input{recommendation}

\ifnum\paperversion=0
\else
\section*{Acknowledgements}

This work is supported by Deutsche Forschungsgemeinschaft (DFG, German Research Foundation) as part of the Research and Training Group 2475 ``Cybercrime and
Forensic Computing'' (number 393541319/GRK2475/1-2019) and by the state of Bavaria at the Nuremberg Campus of Technology (NCT).
\fi

% %-------------------------------------------------------------------------------
% \section*{Availability}
% %-------------------------------------------------------------------------------
%
% USENIX program committees give extra points to submissions that are
% backed by artifacts that are publicly available. If you made your code
% or data available, it's worth mentioning this fact in a dedicated
% section.

% --- -----------------------------------------------------------------
% --- The Bibliography.
% --- -----------------------------------------------------------------
% Add the option --min-crossrefs=1000 to bibtex to avoid having conferences in the bibliography.
% Yes this works for every style
\printbibliography

% --- -----------------------------------------------------------------
% --- The Appendix.
% --- Maybe some appendices shouldn't appear in every version
% --- -----------------------------------------------------------------

 \begin{appendix}

   \input{app-entropy}
    \ifnum\paperversion=1

    \fi

    \ifnum\paperversion=2

    \fi

 \end{appendix}

\end{document}

%% file: 0-packages.tex
\usepackage{array}
\usepackage{booktabs}
\usepackage{centernot}
\usepackage[
	lambda,
	operators,
	advantage,
	sets,
	adversary,
	landau,
	probability,
	notions,
	logic,
	ff,
	mm,
	primitives,
	events,
	complexity,
	asymptotics,
	keys]{cryptocode}
\usepackage[T1]{fontenc}
\usepackage{graphicx}
\usepackage{framed}
\usepackage{mdframed}
\usepackage[utf8]{inputenc}
\usepackage{csquotes}
\usepackage{microtype}
\usepackage{multirow}
\usepackage{paralist}
\usepackage{printlen}
\usepackage{tikz}
\usetikzlibrary{matrix,shapes,arrows,arrows.meta,positioning,chains,calc}
\usepackage{wrapfig}
\usepackage{url}
\usepackage{xcolor}
\usepackage{xspace}
\usepackage[colorinlistoftodos,textsize=small]{todonotes}
\ifnum\style=5
\else
	\usepackage[hidelinks]{hyperref}
\fi
\usepackage{aliascnt}
\usepackage[capitalize,noabbrev,nameinlink]{cleveref}
\usepackage{rotating}
\usepackage{subcaption}
\usepackage{amssymb}
\usepackage{dashbox}

%% file: 0-macros.tex
\def\rvG{G}

\newcommand{\PrG}[1]{\Pr[G\gets\mathcal{G}]{#1}}

\newcommand{\core}[1]{\mathsf{Core}\left(#1\right)}
\newcommand{\regtag}{\mathtt{reg}}
\newcommand{\bintag}{\mathtt{bin}}
\newcommand{\Samp}{\mathsf{Samp}}
\newcommand{\RegSamp}{\mathsf{RegSamp}}
\newcommand{\BinSamp}{\mathsf{BinSamp}}
\newcommand{\SC}{\Gamma}
\newcommand{\id}{\ensuremath{\mathsf{id}}\xspace}
\newcommand{\ID}[2][]{\mathsf{id}_{#1}(#2)}

\DeclareMathOperator*{\sfprob}{\mathsf{Pr}}
\renewcommand*{\Pr}[2][]{\ensuremath{\sfprob_{#1}\left[#2\right]}}

\def\ImportFromMnSymbol#1{%
	\DeclareFontFamily{U} {MnSymbol#1}{}
	\DeclareFontShape{U}{MnSymbol#1}{m}{n}{
		<-6> MnSymbol#15
		<6-7> MnSymbol#16
		<7-8> MnSymbol#17
		<8-9> MnSymbol#18
		<9-10> MnSymbol#19
		<10-12> MnSymbol#110
		<12-> MnSymbol#112}{}
	\DeclareFontShape{U}{MnSymbol#1}{b}{n}{
		<-6> MnSymbol#1-Bold5
		<6-7> MnSymbol#1-Bold6
		<7-8> MnSymbol#1-Bold7
		<8-9> MnSymbol#1-Bold8
		<9-10> MnSymbol#1-Bold9
		<10-12> MnSymbol#1-Bold10
		<12-> MnSymbol#1-Bold12}{}
	\DeclareSymbolFont{MnSy#1} {U} {MnSymbol#1}{m}{n}
}
\newcommand\DeclareMnSymbol[4]{\DeclareMathSymbol{#1}{#2}{MnSy#3}{#4}}

\ImportFromMnSymbol{C}
\DeclareMnSymbol{\smalltriangle}{\mathbin}{C}{73}
\def\upp{\smalltriangle}

\makeatletter
\DeclareOldFontCommand{\bf}{\normalfont\bfseries}{\mathbf}
\DeclareOldFontCommand{\it}{\normalfont\itshape}{\mathit}
\DeclareOldFontCommand{\rm}{\normalfont\rmfamily}{\mathrm}
\DeclareOldFontCommand{\sc}{\normalfont\scshape}{\@nomath\sc}
\DeclareOldFontCommand{\sf}{\normalfont\sffamily}{\mathsf}
\DeclareOldFontCommand{\sl}{\normalfont\slshape}{\@nomath\sl}
\DeclareOldFontCommand{\tt}{\normalfont\ttfamily}{\mathtt}
\makeatother

\def\gkeywords{}
\def\gabstract{}

\NewDocumentCommand\allkeywords{m}{\def\gkeywords {#1}}
\NewDocumentCommand\allabstracts{m}{\def\gabstract{#1}}

\newcommand\makealltitles{
\ifnum\style=\sigalternate
    {
        \begin{abstract}{\gabstract}\end{abstract} \maketitle%
    }
\else
  \ifnum\style=\sciendo
      {\begin{abstract}{\gabstract}\end{abstract}
      }\keywords{\gkeywords}
      \maketitle
  \else
    {\maketitle%
    \ifnum\style=\lncs
      {\begin{abstract}{\gabstract \keywords{\gkeywords}}\end{abstract}}
    \else
      {\begin{abstract}{\gabstract}\end{abstract}
        \ifnum\style=\ieeetr
        \begin{IEEEkeywords}
          {\gkeywords}
        \end{IEEEkeywords}
        \fi
      }
    \fi
    }
  \fi
\fi
}

\mathchardef\mhyphen="2D

\newcommand{\eg}{{e.g.}\xspace}
\newcommand{\etal}{{\it et~al.}\xspace}

\newcommand{\ie}{{i.e.}\xspace}
\newcommand{\ignore}[1]{}

\makeatletter % for auto indention without end...
\newcommand\pcend{\@pc@decreaseindent}
\makeatother

\newcommand{\tuple}[1]{\left( #1 \right)}

\newcommand{\cG}{\ensuremath{\mathcal{G}}\xspace} % Class of functions G

\newcommand{\cX}{\ensuremath{\mathcal{X}}\xspace}
\newcommand{\cY}{\ensuremath{\mathcal{Y}}\xspace}
\newcommand{\cR}{\ensuremath{\mathcal{R}}\xspace}

\let\oldprocedure\procedure
\renewcommand\procedure[3][]{\hspace{0pt}\oldprocedure[#1]{#2}{#3}}

\unless\ifnum\style=\sigalternate

\makeatletter

\newcommand\grantnum[3][]{#3%
  \def\@tempa{#1}\ifx\@tempa\@empty\else\space(\url{#1})\fi}
\makeatother
\fi

\NewDocumentCommand\inputanon{m}{
\ifnum\paperversion=\anonymous \input{#1} \fi
}

\NewDocumentCommand\inputcam{m}{
\ifnum\paperversion=\cameraready \input{#1} \fi
}

\NewDocumentCommand\inputfull{m}{
\ifnum\paperversion=\fullversion \input{#1} \fi
}

\NewDocumentCommand\newuser{mmmo}
{
	\expandafter\NewDocumentCommand\csname todo#1\endcsname{sO{}m}{
		\IfBooleanTF{##1}
		{\unless\ifnum\showcomments=0{\IfNoValueTF{#4}{\color{#3}}{\color{#4}}\textbf{#2:} ##3}\xspace\fi
    }
		{	\quitvmode% ensures that inline todos after pargraph headings are properly displayed
			\texorpdfstring{\todo[inline,color=#3,##2]{\textbf{#2:} ##3}\xspace}{(TODO: #2: ##3)}%
		}%
	}
}

\renewcommand\epsilon\varepsilon

%% file: 0-abstract.tex
% !TEX root = main.tex
In a ring-signature-based anonymous cryptocurrency, signers of a transaction are hidden among a set of potential signers, called a ring, whose size is much smaller than the number of all users.  The ring-membership relations specified by the sets of transactions thus induce bipartite transaction graphs, whose distribution is in turn induced by the ring sampler underlying the cryptocurrency.

Since efficient graph analysis could be performed on transaction graphs to potentially deanonymise signers, it is crucial to understand the resistance of (the transaction graphs induced by) a ring sampler against graph analysis. Of particular interest is the class of partitioning ring samplers. Although previous works showed that they provide almost optimal local anonymity, their resistance against global, e.g. graph-based, attacks were unclear.

In this work, we analyse transaction graphs induced by partitioning ring samplers. Specifically, we show (partly analytically and partly empirically) that, somewhat surprisingly, by setting the ring size to be at least logarithmic in the number of users, a graph-analysing adversary is no better than the one that performs random guessing in deanonymisation up to constant factor of 2.

%% file: intro.tex
% !TEX root = main.tex

\section{Introduction}

In many anonymous systems, a main cryptographic component for providing anonymity is a linkable ring signature (LRS) scheme~\cite{ACISP:LiuWeiWon04}, which is a signature scheme with a restricted anonymity guarantee.
The goal of this work is to study the resistance of these systems against graph-based deanonymisation attacks. For concreteness, we will use privacy-preserving cryptocurrencies as a running example of anonymous systems based on LRS schemes.
We emphasise, however, that the techniques introduced in this work are also directly applicable to other applications of LRS, e.g. anonymous voting~\cite{ISC:YLSNSR18,DBLP:conf/voteid/PatachiS17}\footnote{Although these schemes as described include all legitimate voters in rings, using smaller rings is more efficient. The analyses provided in this work allow designers to make an informed decision of how smaller ring sizes could be chosen.}.
% {\color{OliveGreen}
%   In particular, our analysis directly applies to voting systems such as \cite{ISC:YLSNSR18,DBLP:conf/voteid/PatachiS17} which are based on ring signatures.
%   While the schemes are described using ring signatures which capture all legitimate voters, using smaller rings is desirable for efficiency reasons there as well and our analysis allows doing so while still maintaining a thorough security analysis.
% }

\subsection{Linkable Ring Signatures}

We begin by recalling the basics of LRS schemes.
To sign a message $\mu$, \eg a transaction in a cryptocurrency, the signer first samples a \emph{ring} $r$, \ie a set consisting of (the
%{\color{OliveGreen}
  public keys
%}
of) the signer itself and decoys, using an external algorithm known as a \emph{ring sampler}, then uses the LRS scheme to produce a signature $\sigma$. The tuple $(r,\mu,\sigma)$ is communicated to the verifiers, \eg by publishing it on the blockchain in the context of cryptocurrencies.
%{\color{OliveGreen}
In applications, it is common for a human user to own many pairs of public and secret keys.
Nevertheless, to simplify terminologies, we will refer to a public key as a ``user'' and
%}
use the notation $U$ to refer to the set of users
% {\color{OliveGreen}
(public keys)
% }
where signers belong to and where decoys are sampled from. Depending on the application, the set $U$ could grow over time.

An LRS scheme is \emph{linkable} in the sense that there exists an efficient public algorithm to determine whether any two given signatures are generated by the same signer, \ie using the same secret key. Applications of LRS schemes often employ a ``single-sign verification rule'' which only accepts new signatures which are not linked to any previously accepted ones, so that each user can only perform certain anonymous action once.
For example, such anonymous action could be spending a coin in a cryptocurrency, casting a vote in an anonymous voting system, authenticating and redeeming scores in an anonymous credential system, etc.
In general, the single-sign rule ensures that, at any time, each signer in the set of users has at most one signature that is accepted by the verifiers.

% On one hand, the signer is hidden within a group of potential signers, called a \emph{ring}, sampled from the \emph{universe} of signers using an external algorithm called a ring sampler.
% linkable ring signatures
% On the other hand, signatures issued by the same signer are \emph{linkable}, which effectively restricts each signer to only sign once to remain anonymous.

The \emph{anonymity} of an LRS scheme guarantees that the tuple $(r, \mu, \sigma)$ leaks no more information (computationally) about the signer creating $\sigma$ than what is leaked by the ring $r$ sampled by the ring sampler.
% \textcolor{OliveGreen}{
% In other words, our notions of signers and their anonymity concern the secret key level instead of the human user level. Note that the former is stronger since, if one cannot link transactions to keys, one also cannot link them to humans. The former is also more general since the case of each user owning exactly one key is captured as a special case where there is a one-to-one correspondence between the two levels.}
% \footnote{
% \textcolor{OliveGreen}{Our notion of anonymity is stronger than the usual notion considered for Bitcoin. Concretely, Bitcoin (as is) is vulnerable to considerably stronger deanonymisations than those against Monero, even if the signers of all Monero rings are exactly identified. This is because, apart from ring signatures, Monero also features a mechanism known as ``stealth address'' which provides cryptographic receiver anonymity. Perfect receiver anonymity can also be achieved in Bitcoin (at the cost of heavy key management) if fresh receiver addresses are generated for each new transaction.}
% }
Typically, for efficiency reasons, the ring size $|r|$ is much smaller than the number of users $|U|$, making it plausible to deanonymise signers just by observing ring membership relations implied by the set of published rings, regardless of how secure the LRS scheme is.\footnote{
For example, at the time of writing, Monero mandates a ring size of $|r| = 11$ and has a number of  public keys $|U| \geq 16 \times 10^6$.
%{\color{OliveGreen}
  We note that we are considering anonymity at the key level, which is a stronger notion than anonymity at the human user level typically considered for non-anonymous cryptocurrencies such as Bitcoin.
	Indeed, even if all spenders in Monero transactions are deanonymised, receivers would still be cryptographically anonymous due to the ``stealth address'' mechanism.
%}
}
It is therefore important to design ring samplers and choose their parameters in a way that strikes a balance between efficiency and anonymity, which is the central topic of this work.

\subsection{Transaction Graphs}
\label{sec:transaction_graph}

	To understand deanonymisation attacks of the above kind, we model ring membership relations by \emph{transaction graphs}.
	Specifically, consider an application of LRS where, at some point in time, the tuples $\set{(r_j,\mu_j,\sigma_j)}_{r_j \in R}$ are accepted by the verifiers, where $R$ is some set of rings and, for all $r_j \in R$, members of the ring $r_j$ were sampled from $U$. The ring membership relations can be represented by a transaction graph, which is a bipartite graph $G$ with vertex sets $U$ and $R$, and $u_i \in U$ is connected to $r_j \in R$ if user $u_i$ is a member of ring $r_j$. \cref{fig:toytxg} is a toy example of a transaction graph consisting of 3 users and 3 rings.
	%Assuming that all $R_j$'s are sampled using a ring sampler $\Samp$ and the outcome of $\cG^\Samp$ are illustrated in the transaction graph in~\cref{fig:toytxg}. Further assume that the transaction graph $G$ follows a distribution $\cG^\Samp$ induced by $\Samp$ and the randomness used to sample the $R_j$'s.

A transaction graph is guaranteed to have a maximum matching involving the vertex set $R$. Indeed, by the unforgeability of the LRS we can assume that $\sigma_j$ was issued by some signer $u_j \in r_j$ for each $r_j \in R$, and by the linkability of the LRS and the single-sign verification rule we can assume that all $u_j$'s are distinct. This means that the set $\set{(u_j, r_j)}_{r_j \in R}$ is a maximum matching.

A transaction graph could have many maximum matchings, each representing a possible assignment of signatures/rings to signers.
The union of all maximum matchings of $G$ is known as the Dulmage-Mendelsohn (DM) decomposition~\cite{CJM:DulMen58} or simply the \emph{core} $\core{G}$ (in the sense of DM), and can be computed in linear time given $G$~\cite{TCS:Tassa12}.
If an edge $(u_i,r_j) \in G$ does not belong to any maximum matching, \ie $(u_i,r_j) \notin \core{G}$, then user $u_i$ cannot have been the signer creating $\sigma_j$.
Consequently, the signature-signer assignments represented by the edges $G \setminus \core{G}$ can be ruled out given the knowledge of $\core{G}$.
In extreme cases, where a user $u_i$ is connected to only a single ring $r_j$ in $\core{G}$, the user $u_i$ is considered completely deanonymised.

\input{fig_toytxg}

Referring to the example in~\cref{fig:toytxg}, upon knowing that the only maximum matching is $\{(u_j,r_j): j=1,2,3\}$, in other words $G\neq\core{G}$ and $G\setminus \core{G} = \{(u_1,r_3),(u_2,r_3)\}$, all three signers can be deanonymised. Note how user $u_3$ is deanonymised due to the memberships of the other two rings, although its ring consists of three members. We see that the anonymity of a signer does not only depend on its own ring, but also on the other rings. A global view on the transaction graph is thus required to properly assess the anonymity of signers.

Another richer and more realistic example is given later in~\cref{fig:induced_digraph} (\cpageref{fig:induced_digraph}), which shows a transaction graph $G$ with 8 users and 7 rings, and all rings consist of more than one member. On computing $\core{G}$, we see that 4 out of the 19 potential signature-signer assignments can be ruled out, and one of the signers, namely user 4, can be completely deanonymised.

\subsection{Graph-based Deanonymisation Attacks}

Generalising the attack illustrated in~\cref{fig:toytxg,fig:induced_digraph}, we consider graph-based deanonymisation attacks, where an adversary attempts to identify the signer who sampled $r_{j^*} \in R$, for some $j^*$ chosen by the adversary, given only a transaction graph $G$ representing all rings $R$. In particular, we consider adversaries which do not attempt to break the LRS scheme and which do not have knowledge about the signing probabilities of the signers. The former is easily justified since the LRS scheme is supposedly cryptographically secure. The latter is sensible when the signing probabilities are (close to) uniform by heuristics, \eg when using a partitioning ring sampler to be discussed in~\cref{sec:intro_part}. Finally, our security model capturing untargeted attacks is strong, since if an adversary is successful in a targeted attack, then it is also successful in an untargeted one.

A trivial attack strategy is to choose the smallest ring $r_{j^*} \in R$ and output one of the ring members uniformly at random, which has success probability of exactly $1/|r_{j^*}|$.
We therefore want to upper-bound the success probability of any graph-analysing adversary such that it is not much greater than that of the trivial strategy.
Our strategy is to show that the success probability of an adversary is at most $\Pr{G \neq \core{G}}$ greater than that of the trivial strategy mentioned above, where $G$ is a transaction graph induced by the ring sampler of interest, i.e. the best non-trivial strategy that an adversary could use is to perform DM decomposition.

Although DM decomposition is a well-known tool in graph theory, the technique seems to be adopted only recently to analyse anonymous cryptocurrencies~\cite{EPRINT:Vijayakumaran21}, where it is shown that the analytical deanonymisation attack on Monero based on DM decomposition is at least as effective as existing attacks~\cite{ESORICS:KFTS17,PoPETS:MSHLHSHHMNC18,FC:YAYYXL19} of the same nature.
Indeed, this is as expected since all existing attacks are graph-based and the signature-signer assignments ruled out by these attacks could also be found by DM decomposition.
However, a broader understanding of graph-based attack on ring samplers appears to be lacking.
In particular, the previous examples of attack lead us to the question:
\emph{How should rings be chosen such that the success probability of a graph-based attack can be upper-bounded?}

\subsection{Partitioning Samplers}\label{sec:intro_part}

Of particular interest are the partitioning samplers~\cite{PoPETs:RELSY21}, which first publicly partition the set of users into chunks, randomly choose $k$ decoys from the chunk that the signer belongs to, and output the set which contains the signer and the $k$ decoys as the ring.  Assuming that for each chunk the signing probabilities of the signers in the chunk are close to each other, a partitioning sampler provides near-optimal \emph{local} anonymity according to an entropy-based measure~\cite{PoPETs:RELSY21}, which we discuss further in both~\cref{sec:related-metrics} and~\cref{app:entropy}. Furthermore, in the extreme case that all chunks of the partition are of size $k+1$ -- equal to the ring size -- then the induced transaction graph $G$ simply consists of disjoint $(k+1)$-bicliques and $G = \core{G}$ trivially.
Despite having these features, little is known about the \emph{global} anonymity, \eg the resistance against graph analysis, of partitioning samplers for general chunk sizes.

\subsection{Our Contributions}

In this work, we study the resistance of ring samplers against graph-based deanonymisation attacks.
More precisely, let $\cG^\Samp$ be the distribution of transaction graphs induced by a ring sampler $\Samp$.
We derive an upper bound of $\Pr[G \gets \cG^\Samp]{\rvG \neq \core{\rvG}}$ by relating the event $G \neq \core{G}$ to that of certain digraphs induced by $G$ being not strongly connected.
In case $\Samp$ is a partitioning sampler, we show that this probability likely upper-bounds the advantage of any adversary performing graph-based deanonymisation attacks.

Specifically, assuming two conjectures on certain distributions of random directed graphs (digraphs) which we support by providing empirical evidence, we show that if the number of decoys $k$ of a partitioning sampler is set to
\[
	k \geq \ln (2 \cdot |U|) + \sqrt{2 \ln (2 \cdot |U|)},
\]
then $\Pr[G \gets \cG^\Samp]{\rvG \neq \core{\rvG}} \leq \frac{1}{k+1}$. In other words, a graph-analysing attack is at most twice as successful as a trivial attack does.

Since graph-based attacks threaten all decoy-based anonymous systems, such as coin-mixing, mix-nets, and voting, not limited to LRS-based cryptocurrencies, our result is broadly applicable: It serves as a guideline for choosing parameters for all such systems to avoid graph-based deanonymisation attacks.

\subsection{Technical Overview}

%This work targets anonymous cryptocurrencies which are based on a linkable ring signature scheme (RSC), and our goal is to upper-bound the advantage of an adversary gained from performing graph analysis through upper-bounding the probability of $G\neq\core{G}$. Specific focus is paid to the case where a partitioning sampler~\cite{PoPETs:RELSY21} is deployed. In the following we first provide a brief summary on RSC, followed by a high level walk-through of our results, which are depicted by the sequence of inequalities in~\cref{fig:outline}.

% Our results are obtained through a sequence of steps
% depicted in~\cref{fig:outline},
% which we summarise below.

For the ease of reading the technical sections, we provide a high-level overview below.

\subsubsection{Transaction Graphs and Induced Digraphs}
The central objects studied in this work are transaction graphs and their induced digraphs, which are formally defined in \cref{sec:graphs}.
% We focus on the case of single-signer distributions.
As described in \cref{sec:transaction_graph}, a transaction graph is a bipartite graph $G$ with vertex sets $(U,R)$ and edges $E$. %The sets $N$ and $M$ are interpreted as the index sets for the signers and the rings respectively, where the index of a ring is the index of the signer who creates it. The edges $E$ represents the ring membership relations, \ie $(i,j) \in E$ if signer $i$ belongs to ring $j$. When $N = M$, we say that the graph $G$ is balanced.
%As an illustration, in the toy example in \cref{fig:..}, the universe consists of 3 signers and all have generated a ring. To read the graph, for example, edges $(1,1)$ and $(2,1)$ represent that signers 1 and 2 are members of the ring generated by signer 1, \ie signer 2 is the decoy.
For any transaction graph $G$, suppose without loss of generality that $M = \set{(u_j,r_j)}_{j=1}^m$ is a maximum matching in $G$. We can define its induced digraph $\ID{G}$ such that $(i,j)$ is an edge in $\ID{G}$ whenever $(u_i,r_j)$ is an edge in $G$ and $i \neq j$. We use $\vec{G} \in \SC$ to denote that $\vec{G}$ is strongly connected.

\subsubsection{Modelling Graph-based Deanonymisation}

To model the security of ring samplers against graph-based deanonymisation attacks, in~\cref{sec:ring_samp_defs}, we first formalise the notion of ring-sampler-induced transaction graphs, then model the security by designing a security experiment.

For any ring sampler $\Samp$ and any number of signatures $m \leq |U|$ , we define the induced transaction graphs sampler $\cG^{\Samp}$ which inputs $(U,1^m)$ and outputs a tuple $(G,M)$ where $G = (U,R,E)$ with $|R| = m$ is a transaction graph induced by $\Samp$ and $M$ is a maximum matching in $G$.
%The sampler $\cG^{\Samp}$ samples $(G,M)$ by running $r_j \gets \Samp(U,s_j)$ for a random $m$-subset of signers $\set{s_1,\ldots,s_m} \subseteq U$, resulting in a transaction graph $G$, and setting $M = \set{(s_j,r_j)}_{j=1}^m$. When the maximum matching $M$ is not of interest, we omit it and write $G \gets \cG^{\Samp}(U,m)$.

We say that $\Samp$ is $\epsilon$-secure against graph-based deanonymisation attacks if no adversary, when given a transaction graph $G$ where $(G,M) \gets \cG^{\Samp}(U,1^{|U|})$, could find an edge in $M$, i.e. a signer-ring assignment, with probability more than $\epsilon$. The setting of $m = |U|$ in $\cG^{\Samp}(U,1^{|U|})$ is without loss of generality due to~\cref{thm:balance_is_upper_bound}, to be explained in~\cref{sec:tech_overview_regular}. While the focus of this work is on passive adversaries, we also define a more general notion of security against active adversaries who compromise an admissible subset of users.
The generalised notion captures the so-called ``black marble attacks''~\cite{MRL-0001,MRL-0004,8456034} in the literature.

A trivial strategy of the adversary is to pick the smallest ring $r^*$ in $G$ and output a random edge connecting such ring, with success probability $1/|r^*|$. Therefore, a sampler $\Samp$ which outputs rings of a fixed size $k+1$ cannot be $\epsilon$-secure for $\epsilon < \frac{1}{k+1}$.
Intuitively, the trivial strategy is also the best strategy for the adversary in case $G = \core{G}$, which we prove to be the case for partitioning samplers in~\cref{sec:interpretation}.
Hence, to upper-bound the success probability of any adversary against $\Samp$, it suffices to upper-bound the probability that $G \neq \core{G}$ for transaction graphs $G$ induced by $\Samp$.

\subsubsection{Problem Reduction}
Our first step for upper-bounding $\Pr{G\neq\core{G}}$, carried out in~\cref{sec:reduction}, is to reduce the problem about $\core{\rvG}$ of a transaction graph $G$, a somewhat unwieldy object, to a simpler problem about the induced digraphs of the subgraphs of $G$. Although the results in~\cref{sec:reduction} hold for general transaction graphs, they are motivated by the observation that the transaction graphs induced by partitioning samplers could be partitioned into subgraphs whose induced digraphs follow some simple-to-describe distributions. The reduction is summarised by \cref{thm:core}, which states that
$\Pr{G\neq\core{G}}$ is upper-bounded by a sum of probabilities of some induced digraphs being not strongly connected.
%\begin{align*}
%	\PrG{\rvG \neq \core{\rvG}}
%	\leq \sum_{i=1}^\ell \Pr[G_i \gets \mathcal{G}_i]{\ID{G^\upp_i} \notin \SC},
%\end{align*}
%where $\ID{G^\upp_i}$'s are the induced digraphs of some subgraphs $G^\upp_i$'s of $G$, and $\SC$ is the set of all strongly connected digraphs.
%This theorem says that, the probability of $G\neq\core{G}$ is upper-bounded by a sum of probabilities of some digraphs being not strongly connected.

%This theorem is proved in two main steps.
%First, in~\cref{cor:prodH}, we show that
%\begin{align*}
%\PrG{\rvG \neq \core{\rvG}}
%\leq \sum_{C \in P} \Pr[G_C \gets \mathcal{G}_C]{G_C \neq \core{G_C}}
%\end{align*}
%if the set of distributions $\set{\cG_C}_{C \in P}$ is a partition of the distribution $\cG$, in the sense that sampling from $\cG$ is equivalent to first sampling from each of $\cG_i$ and then taking the union.
%This allows us to consider the cores $\core{G_C}$ of the subgraphs $G_C$ of $G$.
%
%Next, in~\cref{lem:sc}, we show that
%\[
%\PrG{\rvG \neq \core{\rvG}} \leq \PrG{\ID{G} \notin \SC}.
%\]
%This allows us to further reduce the problem of $\core{G_C}$ to that of the strong connectivity of $\ID{G_C}$.

\subsubsection{Regular Partitioning Samplers}\label{sec:tech_overview_regular}
In \cref{sec:partitioning_samplers}, we move on to identify the transaction graphs induced by a partitioning sampler and their induced digraphs.
Intuitively, the more information that is available to an adversary, the better it could perform in deanonymisation attacks, \eg through graph analysis.
%\textcolor{OliveGreen}{
Indeed, we show in~\cref{thm:balance_is_upper_bound} that for any number of signers $m \leq |U|$, the probability of $G \neq \core{G}$ where $G$ is transaction graph sampled by a ring sampler is upper-bounded by that when $m =|U|$. This allows us to consider simply the latter case, which corresponds to that all users have signed.%}
%\begin{align*}
%	&\Pr[G \gets \cG^\Samp(U,m)]{G \neq \core{G}} \\
%	\leq &\Pr[H \gets \cG^\Samp(U,|U|)]{H \neq \core{H}}.
%\end{align*}
%We remark that this result holds for any ring sampler $\Samp$.

Next, we focus on the partitioning ring samplers proposed in~\cite{PoPETs:RELSY21}, denote by $\Samp = \RegSamp[P,k]$, which are parametrised by a partition $P$ of $U$ and a number of decoys $k$. The notation $\RegSamp$ stands for regular partitioning sampler, whose naming shall become clear shortly below. On input a signer $s$, $\RegSamp[P,k]$ locates the chunk $C \in P$ which contains the signer $s$, samples a uniformly random $(k+1)$-subset $r$ of $C$ conditioning on $s \in R$, and outputs $r$ as the ring.

A convenient property of a partitioning sampler $\RegSamp[P,k]$ is that its distribution of induced transaction graphs can be naturally partitioned.
Going through the reduction established in~\cref{sec:reduction}, we observe that the induced digraphs of each chunk in the partition follows the uniform distribution over all $k$-in-degree regular (hence the notation $\RegSamp$) digraphs with $n = |C|$ vertices, denoted by $\vec{\cG}^{\regtag}_{k,n}$.
This, however, presents a challenge to our goal of upper-bounding the probability of $G\neq\core{G}$, since the distributions $\vec{\cG}^{\regtag}_{k,n}$ do not appear to be well-studied in random graph theory.

% \paragraph*{Binomial Partitioning Samplers.}
\subsubsection{Conjectures and Empirical Evidences}

% Towards circumventing the above problem, we propose and analyse a new variant of partitioning samplers called the binomial partitioning samplers, whose naming will become clear shortly below.
% Similar to regular partitioning samplers $\RegSamp[P, k]$, binomial partitioning samplers $\BinSamp[P, p]$ are parametrised by a partition $P$ of $U$ and a probability $p$ which determines the expected ring size $p|C| + 1$ for each chunk $C \in P$. On input a signer (index) $j$, $\BinSamp[P,p]$ locates the chunk $C \in P$ which contains the signer $j$, initialises $R = \set{j}$, includes each member of $C \setminus \set{j}$ into $R$ with probability $p$, and outputs $R$ as the ring.
%
% As in the regular case, the distribution $\cG^{\BinSamp[P,p]}_{U,U}$ also features convenient properties. First, it is clear that $\cG^{\BinSamp[P,p]}_{U,U}$ can be naturally partitioned into $\set{\cG^{\BinSamp[P,p]}_{C,C}}_{C \in P}$. Furthermore, we observe that $\ID{\cG^{\BinSamp[P,p]}_{C,C}}$ follows the distribution $\vec{\cG}^{\bintag}_{p,|C|}$, the distribution over digraphs with vertex set $C$ where each possible edge appears with probability $p$, up to a renaming of the nodes. Note that the in-degree (and the out-degree) of each node in $\vec{G} \gets \vec{\cG}^{\bintag}_{p,n}$ follows a binomial distribution (hence the notation $\BinSamp$) for $n \in \NN$.

Towards circumventing the above problem, in \cref{sec:conjectures}, we turn our attention to the distribution $\vec{\cG}^{\bintag}_{p,n}$ over digraphs with $n$ vertices where each possible edge appears with probability $p$, with the intuition that the strong connectivity of $\vec{\cG}^{\regtag}_{k,n}$ could be estimated by that of $\vec{\cG}^{\bintag}_{p,n}$ for appropriately chosen $(k,p)$.\footnote{Similar to how a regular partitioning sampler $\RegSamp[P,k]$ relates to the distribution $\vec{\cG}^{\regtag}_{k,n}$, a ``binomial partitioning sampler'' $\BinSamp[P,p]$ could be constructed and be related to the distribution $\vec{\cG}^{\bintag}_{p,n}$. To avoid distraction, we defer a discussion on this to~\cref{app:binomial_samplers}.}

To relate the two distributions, in~\cref{conj:reg_leq_bin}, we conjecture that
\begin{equation*}
	\label{eqn:bin:reg}
	\Pr[\vec{G} \gets \vec{\cG}^\regtag_{k,n}]{\vec{G} \notin \SC}
	\leq
	\Pr[\vec{G} \gets \vec{\cG}^\bintag_{p,n}]{\vec{G} \notin \SC}
\end{equation*}
when $p = \frac{k}{n-1}$ and therefore the expected in-degree for $\vec{G} \gets \vec{\cG}^\bintag_{p,n}$ is $k$.
This makes sense intuitively when considering the natures of both digraph models.
If the conjecture holds, it allows us to consider the distribution $\vec{\cG}^\bintag_{p,n}$, which is better understood.

Based on the result of Pal{\'a}sti~\cite{palasti1966strong}, the distribution $\vec{\cG}^{\bintag}_{p,n}$ was studied by Graham and Pike~\cite{graham2008note}, who proved the limit of $\Pr[\vec{G} \gets \vec{\cG}^{\bintag}_{p,n}]{\vec{G} \notin \SC}$ under specific choice of $p$.
%for any constant $c \in \RR$, if $p(n) := \frac{\ln n + c}{n}$, then
%\[
%	\lim_{n \to \infty} \Pr[\vec{G} \gets \vec{\cG}^{\bintag}_{p,n}]{\vec{G} \notin \SC} = 1 - e^{-2e^{-c}}.
%\]
%Following the above discussion, we face two obstacles.
%First, little was established about the in-degree regular random digraphs $\vec{\cG}^{\regtag}_{k,n}$, and analysing them turns out to be difficult due to its combinatorial nature.
%Second, even for the closely related binomial random digraphs $\vec{\cG}^{\bintag}_{p,n}$, only asymptotic results were known.
Using this result, in~\cref{conj:bin_upper_bound}, we propose our second conjecture, which states that
\begin{equation*}
  \label{eqn:lim:bin}
	\Pr[\vec{G} \gets \vec{\cG}^\bintag_{p,n}]{\vec{G} \notin \SC}
	\leq
	1 - e^{-2e^{\ln n - pn}},
\end{equation*}
where the expression on the right hand side is heuristically obtained from the result of \cite{graham2008note}. %by plugging $c = \ln n - pn$ back into the limit.

Assuming both conjectures and combining all previous results, we conclude a closed-form upper bound for $\Pr{G \neq \core{G}}$.
Although we are unable to prove the conjectures, in~\cref{sec:exp}, we provide empirical evidences that they indeed seem to hold, at least for parameters of interest in the context of cryptocurrencies.
In particular, we sampled $8000$ random graphs according to either distribution in order to estimate the actual probabilities.
We observe that the conjectured inequalities hold for all tested values of $k$ and $n$.
%Interestingly, while the first inequality appears to be tighter when $n$ decreases, the second inequality appears to be tighter when $n$ increases.
% , we note particularly that for $n\geq256$ the upper bound in \cref{eqn:lim:bin} is quite tight as can be seen in \cref{fig:plot:universes}.
% The same can be said for $k$ being as small as 7 (see \cref{fig:plot:rings}).
% In both cases the bound from \cref{eqn:bin:reg} is less tight but also holds.

\subsubsection{Provably Secure Ring Samplers}
Putting everything together, in~\cref{sec:interpretation}, we first show that $\RegSamp[P,k]$ is $\epsilon$-secure for
\[
	\epsilon \leq \Pr{G \neq \core{G}} + \frac{1}{k+1}
\]
where $G$ is a random transaction graph induced by $\RegSamp[P,k]$.
Together with other established results, we prove that
%\[
%	\Pr{G \neq \core{G}} \leq |P| \tuple{1 - e^{-2e^{\ln n - k}}},
%\]
%where $n = \max_{C\in P}|C|$ is the maximum chunk size.
%By applying a bound of Chatzigeorgiou~\cite{journals/icl/Chatzigeorgiou13}, we see that
if
\[k \geq \ln (2 \cdot |U|) + \sqrt{2 \ln (2 \cdot |U|)} \]
then
%\[
%	\Pr{G \neq \core{G}} \leq \frac{1}{k+1}
%\]
%and consequently
$\RegSamp[P,k]$ is $\epsilon$-secure for $\epsilon \leq \frac{2}{k+1}$.
In other words, for this parameter choice, no graph-analysing adversary is likely to perform better than random guessing up to constant factor of 2.

Finally, we conclude our work by discussing the security of $\RegSamp[P,k]$ against active graph-based attacks.

% Since a naive adversary who simply guesses randomly has a success probability of $\frac{1}{k'} = \frac{1}{k+1}$, by setting $k$ so that $\adv(k) \leq \frac{1}{k'}$, no graph-analysing adversary could perform better than random guessing up to constant factor of 2.}
% Assuming that the above upper bound indeed holds and all chunks have equal size $n$, it suffices to set the ring size $k'$ such that
% \[k' - \ln k' \geq \ln (2u) +1.\]
% By applying a bound of Chatzigeorgiou~\cite{journals/icl/Chatzigeorgiou13}, we see that it suffices to set
% \[k \geq \ln (2u) + \sqrt{2 \ln (2u)} .\]

\subsection{Related Work}\label{sec:related}

% There has been work in different directions using transaction graphs.
%
% \todoC*{
%   In the following, we are reviewing other works related to this paper.
%   In particular, we discuss both heuristic graph based attacks which motivate that a thorough handling of the topic is called for
%   as well as recent attempts to provide analytic anonymity metrics related to the setting of ring signatures.
%   Finally we will discuss the state of research concerning random graphs, the tool we employ to derive our result.
% }
%
% \todoI*{Our work is closely related to a number of distinct topics, including graph-based attacks on anonymous transactions, anonymity measures for ring signatures and random graph theory.
% We give a brief summary on some of the established results in these areas.}

We conclude the introduction by discussing related works in the areas of graph-based deanonymisation attacks, anonymity metrics, and random graph theory.

\subsubsection{Graph-Based Attacks}

In recent years, numerous works~\cite{ESORICS:KFTS17,PoPETS:MSHLHSHHMNC18,FC:YAYYXL19} demonstrated that, by reducing the ring membership relations represented by the transaction graph of Monero, it is possible to completely deanonymise signers of certain transactions. These attacks commonly rely on
% exploited the structure of transaction graphs structure directly to deanonymize single signers.
% As a basis they use
the fact that, in an early version of Monero,
% allowed a ring size of 1, \ie only the actual signer was the ring.
it was not mandatory for a signer to include decoys in a transaction.
If such a signer A is chosen as a decoy in a ring sampled by another signer B,
% it cannot be the actual spender in this other ring as otherwise double spending would be possible.
the possibility of A being the real signer of the transaction of B can be ruled out easily by the ring membership relation reduction, thereby reducing the anonymity of B.
% This reduces the anonymity of a signer and is used in an additional cascade attack, which does several steps of the aforementioned anonymity reduction.
This anonymity reduction effect can be propagated to another signer C if it chooses B as a decoy in its ring, causing a chain reaction.

Recently, \citeauthor{EPRINT:Vijayakumaran21} \cite{EPRINT:Vijayakumaran21} proposed to use DM decomposition for deanonymising Monero signers, and showed that this is as effective as the prior methods~\cite{ESORICS:KFTS17,PoPETS:MSHLHSHHMNC18,FC:YAYYXL19}.
Indeed, these prior attacks can be seen as finding certain subsets of edges not being in $\core{G}$ for a transaction graph $G$, and are therefore subsumed by DM decomposition which computes the entirety of $\core{G}$.

We remark that the aforementioned works mainly measure the effectiveness of an attack by counting the number of completely deanonymised signers, focusing little on partial deanonymisation.
%  but still newer transactions can be barely deanonymized.
% While this seems to allow the conclusion that the current Monero sampler is sufficient, it waives the problem of partial deanonymization, \ie if the anonymity is not reduced completely but \eg to 2 out of 20 possible spenders.
In contrast,
% our paper aims to reduce the possible anonymity loss even from $k$ out of $k$ to $k-1$ out of $k$ to zero.
the goal of this work is to upper-bound the probability of any partial deanonymisation.

\subsubsection{Anonymity Metrics}
\label{sec:related-metrics}

% \citeauthor{CSF:YuAuVer19} \cite{CSF:YuAuVer19} used a graph-theoretical approach to provide an anonymity measure for transaction graphs by considering the number of perfect matchings for a certain type of graphs, which corresponds to the permanent of the graph which is known to be hard in this particular context. They further evaluate existing attacks and suggest a partitioning sampler.

\citeauthor{CSF:YuAuVer19} \cite{CSF:YuAuVer19} measured the anonymity of a transaction graph using the number of perfect matchings, which is unfortunately \#P-complete to compute. They also evaluated existing attacks and suggested a partitioning sampler which can be seen as a special case of those proposed in~\cite{PoPETs:RELSY21} and discussed below.

Beyond graph analysis, general deanonymisation attacks could take the signing probabilities of different signers into consideration. In this setting, Ronge~\etal~\cite{PoPETs:RELSY21} proposed to model the anonymity provided by a ring sampler by the min-entropy $H_\infty(s|r)$ of the signer $s$ conditioning on the ring $r$ sampled by the signer $s$.
According to this anonymity measure, the authors also proved that (regular) partitioning samplers are close to optimal assuming that the distribution of signing probabilities in each chunk is close to uniform.
The formal definition of this anonymity measure and the corresponding optimality result on (regular) partitioning samplers are recalled in~\cref{app:entropy}.
% These partitioning samplers also subsume the one suggested in \cite{CSF:YuAuVer19}.
%For the sake of completeness, we analyse the binomial partitioning samplers with the same anonymity measure and show in \cref{thm:entropy} that the two partitioning samplers indeed have the same level of near-optimal anonymity.

A major shortcoming of the anonymity measure of Ronge~\etal~\cite{PoPETs:RELSY21}, however, is that it only captures the \emph{local} anonymity of a single signer given a single ring. In particular, it does not capture \emph{global} attacks such as those based on DM decomposition. Although extensions to the global setting were discussed, it is unclear whether the extended measures are efficiently computable.

We remark, however, that although graph analysis informs us about the anonymity of a ring sampler in the \emph{global} sense, it disregards the distribution of signing probabilities. Consequently, a ring sampler (\eg the uniform sampler) that behaves well under graph analysis might achieve low anonymity according to the entropy-based measure.
We therefore view the two approaches as being complementary with each other.

\subsubsection{Random (Di)graph Connectivity}

Numerous results have been established for the connectivity problem of random (undirected) graphs.
Erd{\"o}s and R{\'e}nyi \cite{erdHos1959random} proved the asymptotic probability of a uniform random graph\footnote{A uniform random graph is a graph that is uniformly sampled from the set of all graphs with a fixed vertex set with certain fixed number of edges.
A uniform random digraph is defined analogously.} being connected. {\L}uczak \cite{luczak1987equivalence} extended the result to binomial random graphs.
Gilbert \cite{gilbert1959random} gave both upper and lower bounds of the probability of a finite binomial random graph being connected.
For $k$-regular random graphs, it is known that such graphs are almost surely connected for $k\geq 2$ \cite{mauldin1981scottish} and almost surely $k$-connected for $k \geq 3$ \cite{bollobas2001random}.

The strong connectivity problem of random digraphs is, however, much worse understood.
Among the existing literature, the majority focuses on infinite graphs. Pal{\'a}sti \cite{palasti1966strong} and Graham and Pike \cite{graham2008note} proved the asymptotic probability of strong connectedness for a uniform random digraph and a binomial random digraph respectively.
Some works studied the asymptotic size of the giant strongly connected component (\eg~\cite{penrose2016strong,pittel2016asymptotic}).
Little seems to be known for finite graphs.
The problem of computing (asymptotically) the probability of a $k$-in(/out)-degree regular random digraph being strongly connected was listed in The Scottish Book \cite{mauldin1981scottish} in 1981, and in its second edition in 2015 this problem remains open.
The reachability problem, which asks the probability that a given node can reach all other nodes in a random digraph, though intuitively simpler than the strong connectivity problem, is proven to be \#P-complete~\cite{provan1983complexity}.

%% file: fig_toytxg.tex
% !TEX root = main.tex

\begin{figure}[hb]
	\centering
    \begin{tikzpicture}[every node/.style={circle,draw,minimum size=.3cm,inner sep=0pt}]
    \foreach \i in {1,...,3}{
    	\node[label=left:$u_{\i}$] (s\i) at (0,-\i) {};
    }
    \foreach \i in {1,...,3}{
    	\node[label=right:$r_{\i}$] (r\i) at (2,-\i) {};
    	\draw[very thick,red] (s\i) -- (r\i);
    }
    \begin{scope}[very thick]
%    	\draw (s1) -- (r2);
    	\draw (s1) -- (r3);
    	\draw (s2) -- (r3);
    \end{scope}
	\node[opacity=0, text opacity=1] (U) at (0,-0.2) {\parbox{2.5em}{\centering \small Users\\$U$}};
	\node[opacity=0, text opacity=1] (R) at (2,-0.2) {\parbox{2.5em}{\centering \small Rings\\$R$}};
    \end{tikzpicture}
	\caption{Toy example of transaction graph. Edges correspond to ring memberships, \eg $(u_1,r_3)$ means user 1 is a member of ring 3. The red edges are the only maximum matching.}
	\label{fig:toytxg}
\end{figure}
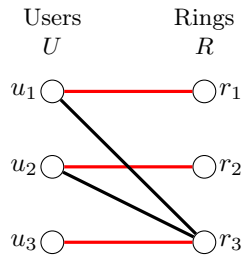

%% file: prelim.tex
% !TEX root = main.tex

% \section{Notation}
\section{Graphs} \label{sec:graphs}

For $n \in \NN$, write $[n] := \set{1,2,\ldots,n}$.
%Let $A, B$ be sets.
%If $A \subseteq B$ and $A$ has $|A| = n$ elements, we write $A \subseteq_n B$.
A partition $P$ of a set $U$ is a set of disjoint subsets of $U$, called chunks, satisfying $\bigcup_{C \in P} C = U$.
We often use $U$ to denote the set of all users, i.e. potential signers.
%and refer to it as the ``universe''.

% \subsection{Graphs}
We assume the general familiarity of the concepts of bipartite graphs and directed graphs (digraphs). In the following, we recall and establish some concepts which are specific to this work.

\subsection{Bipartite Graphs}

A bipartite graph $G = (A,B,E)$ consists of the vertex sets $(A,B)$ (whose elements are also called nodes) and a set $E \subseteq A \times B$ of edges.
Let $G = (A,B,E)$ and $H=(A',B',E')$ be bipartite graphs. We define the following basic operations and relations:
\begin{itemize}
	\item Subgraph: $H$ is a subgraph of $G$, denoted by $H \subseteq G$, if $A' \subseteq A$, $B' \subseteq B$ and $E' \subseteq E$.
	\item Union: $G \cup H := (A \cup A', B \cup B', E \cup E')$.
	\item Intersection: $G \cap H := (A \cap A', B \cap B', E \cap E')$.
	\item Difference: $G \setminus H := (A^-,B^-,E^-) \subseteq G$ where $A^- = A \setminus A'$, $B^- = B \setminus B'$ and $E^- = E \cap (A^- \times B^-)$.
	\item Edge elements: if $e \in E$, we sometimes abuse the notation and write $e \in G$.
\end{itemize}

Our analyses are primarily based on the concept of matchings in bipartite graphs, which we recall below.

\begin{definition}[Matching]
	A matching $M \subseteq E$ in a bipartite graph $G=(A,B,E)$ is a subset of edges such that for all edges $(a,b), (a',b') \in M$, it holds that $a \neq a'$ and $b \neq b'$.
	We say $M$ is a maximum matching, if $|M| \geq |M'|$ for any matching $M'$ in $G$.
%	We say $X$ is a perfect matching, if for any $i \in N$ there exists $(i',j') \in X$ such that $i = i'$, and for any $j \in M$ there exists $(i',j') \in X$ such that $j = j'$.
\end{definition}

\begin{definition}[Core]
	The core of a bipartite graph $G=(A,B,E)$, denoted by $\core{G}=(A,B,E')$, is a subgraph of $G$ where $E' \subseteq E$ is the union of all maximum matchings in $G$.
\end{definition}

The above concept of core is defined in the sense of Dulmage and Mendelsohn~\cite{CJM:DulMen58}. It should not be confused with the core defined with respect to graph homomorphisms. Tassa~\cite{TCS:Tassa12} gave an algorithm for computing $\core{G}$ in time linear in the number of nodes and edges of $G$.

A transaction graph is a bipartite graph $G = (U,R,E)$, where $U$ is a set of users and $R$ is a set of rings\footnote{More precisely, $R$ is a set of ring identifiers. This is to handle cases where different signers sample the same ring.}, such that $|R| \leq |U|$ and there exists at least one maximum matching of size $|R|$.
The edges $E$ capture ring memberships, that is, if user $u_i$ belongs to ring $r_j$, then $(u_i,r_j) \in E$.
The existence of a size-$|R|$ maximum matching captures the assumption that each ring is generated by a distinct signer.
% Without loss of generality we assume that ring $j$ is generated by signer $j$.

\begin{definition}[Transaction Graph]
	A transaction graph $G = (U,R,E)$ is a bipartite graph with a maximum matching $M$ of size $|R| \leq |U|$.
	We say that $G$ is balanced if $|U| = |R|$. Otherwise it is imbalanced.
\end{definition}

By renaming of nodes, we can write $U = \set{u_i}_{i=1}^n$, $R = \set{r_j}_{j=1}^m$, and $M = \set{(u_j, r_j)}_{j=1}^{m}$  for some $m,n \in \NN$ with $n \geq m$ without loss of generality.

\begin{definition}[Upper Graph]
	Let $G=(U,R,E)$ be a transaction graph, where $U = \set{u_i}_{i=1}^n$ and $R = \set{r_j}_{j=1}^m$, and $M = \set{(u_j, r_j)}_{j=1}^{m}$ be a maximum matching in $G$.
	% An upper graph $G' = (U', R, E')$ of $G$ is a balanced transaction subgraph of $G$ satisfying $E' = E \cap (U' \times R)$.
	% The upper graph $G^\upp = (U^\upp, R, E^\upp)$ is a special upper graph of $G$ where $|U^\upp| = |R|$.
	The $M$-upper graph $G^M = (U^M, R, E^M)$ is a balanced transaction subgraph of $G$ where $U^M := \set{u_j}_{j=1}^m$ and $E^M = E \cap (U^M \times R)$.
	We use $G^\upp$ do denote an $M$-upper graph $G^M$ for an arbitrary $M$ chosen deterministically given $G$.
\end{definition}

% A transaction graph $G$ can have multiple upper graphs.
% If $G$ is balanced, then it has only one upper graph which is itself.
The left panel of \cref{fig:induced_digraph} is an example transaction graph $G$ with a maximum matching $M$.
The upper graph $G^M$ is the subgraph of $G$ in the dotted rectangle.

\input{fig_induced_digraph}

To capture transaction graphs induced by partitioning ring samplers, we define the notion of transaction graph partitioning.

\begin{definition}[Transaction Graph Partitioning]
	Let $U$ be a set of signers and $P$ be a partition of $U$.
    Let $G = (U,R,E)$ and $G_C = (C, R_C, E_C)$ be transaction graphs for $C \in P$.
    We say that $\set{G_C}_{C \in P}$ is a partition of $G$ if
    $\set{R_C}_{C \in P}$ and $\set{E_C}_{C \in P}$ are partitions of $R$ and $E$ respectively.

    Generalising, let $\cG$ be a distribution of transaction graphs with vertex sets $(U,R)$.
    We say that $\set{\cG_C}_{C \in P}$ is a partition of $\cG$, if
    $\cG_C$ is a distribution of transaction graphs with vertex sets $(C, R_C)$ for $C \in P$ and $\cG = \bigcup_{C \in P} \cG_C$, \ie sampling from $\cG$ is equivalent to first independently sampling from $\cG_C$ for all $C \in P$ and then taking the union.
	%\footnote{$\cG = \bigcup_{i=1}^\ell \cG_i$ means that sampling from $\cG$ is equivalent to first independently sampling from $\cG_i$ for all $i \in [\ell]$ and then taking the union.}
\end{definition}

Clearly, if $\set{G_C}_{C \in P}$ is a partition of $G$, then the $G_C$'s have disjoint nodes and edges, and $\bigcup_{C \in P} G_C = G$.

\subsection{Digraphs}

A digraph $\vec{G} = (V,E)$ consists of a vertex set $V$ and a set $E \subseteq V^2$ of edges. All digraphs considered in this work are without self-loop and parallel edge. The definitions of basic operations and relations for digraphs are analogous to those for bipartite graphs.

\begin{definition}[Edge Reachability]
	Let $\vec{G} = (V,E)$ be a digraph.
	We say that an edge $e \in E$ is reachable from node $v \in V$ through $\vec{G}$,
	%denoted by $e \stackrel{\vec{G}}{\leftsquigarrow} v$,
	if there exists a directed path $P = \set{(v_i, v_{i+1})}_{i=0}^\ell \subseteq E$ for $v_0=v$ and some $\ell \in \NN$ such that $e \in P$.
	% Generalising, for $U \subseteq V$, we say that $e$ is reachable from $U$ through $\vec{G}$ if $e$ is reachable from $u$ through $\vec{G}$ for some $u \in U$.
	Generalising, if $\vec{H} = (W,F) \subseteq \vec{G}$, we say that $e$ is reachable from $\vec{H}$ through $\vec{G}$ if $e$ is reachable from $w$ through $\vec{G}$ for some $w \in W$. %In this case, $e \stackrel{\vec{G}}{\leftsquigarrow} \vec{H}$.
\end{definition}

The concepts of connectivity and strongly connected components of digraphs will be repeatedly used, their definitions are as follows.

\begin{definition}[Strong and Weak Connectivity]
	A digraph $\vec{G}=(V,E)$ is strongly connected, denoted by $\vec{G} \in \SC$, if %$i \leftsquigarrow j$
	there exists a directed path from $i$ to $j$ for all distinct $i,j\in V$.
	The digraph $\vec{G}$ is weakly connected, if there exists an (undirected) path from $i$ to $j$ for all distinct $i,j\in V$ when disregarding edge orientations.
\end{definition}

\begin{definition}[Strongly Connected Component]
	A strongly connected component (SCC) of a digraph $\vec{G}$ is a subgraph of $\vec{G}$ that is strongly connected, and is maximal with this property -- that is, no further node or edge from $\vec{G}$ can be added to it without breaking its strongly connected property.
\end{definition}

To reduce problems about the cores of transaction graphs to those about digraphs connectivity, we define the notion of induced digraph $\ID{G}$ of a transaction graph $G$.

\begin{definition}[Induced Digraph] \label{def:inducedD}
	Let $G=(U,R,E)$ be a transaction graph, where $U = \set{u_i}_{i=1}^n$ and $R = \set{r_j}_{j=1}^m$, and $M = \set{(u_j, r_j)}_{j=1}^{m}$ be a maximum matching in $G$.
	The $M$-induced digraph of $G$ is defined as $\id_M(G) := ([n], F)$ where
	\(F := \set{(i,j) \in [n]^2: (u_i, r_j) \in E \land i \neq j}\).
	% Given a transaction graph , the induced digraph is $\ID{G} := (U,F)$ where
    % $F = E \setminus \set{(i,i)}_{i \in M}$.
	% In more detail, the induced graphs is constructed as follows:
	% \begin{enumerate}
	% 	\item Enumerate all vertices in $S$ and $R$ separately
	% 	\item Interpret all edges as directed edges from $S$ to $R$ and add them to $F$
	% 	\item Match all vertices in $S$ and $R$ with the same index and remove self-loops
	% \end{enumerate}
	% When the maximum matching is clear from the context or is unimportant, we omit it and write $\id(G)$.
	We use $\id(G)$ to denote an $M$-induced digraph $\id_M(G)$ for an arbitrary $M$ chosen deterministically given $G$.
\end{definition}

In other words, given a maximum matching $M$, if we rename the users and rings so that $u_j \in r_j$ for all $r_j \in R$, the induced digraph is constructed by including an edge from node $i$ to node $j$ if user $u_i$ is a member of ring $r_j$ whenever $i \neq j$.
\cref{fig:induced_digraph} gives an example of an induced digraph $\id(G)$ of a transaction graph $G$.

We further introduce two special types of digraphs which the partitioning samplers will be related to.

\begin{definition}[$k$-In-Degree Regular Digraphs] \label{def:regD}
    Let $k, n \in \NN$ with $k < n$.
    A $k$-in-degree regular digraph is a digraph where all nodes have a fixed in-degree $k$.
    We write $\vec{\cG}^\regtag_{k,n}$ for (the uniform distribution over) the set of all $k$-in-degree regular digraphs with the vertex set $[n]$.
\end{definition}

\begin{definition}[$p$-Binomial Digraphs] \label{def:binD}
    Let $p \in [0,1]$ and $n \in \NN$.
    We write $\vec{\cG}^\bintag_{p,n}$ for the distribution obtained by (uniformly) sampling a digraph $\vec{G}$ with the vertex set $[n]$ such that each of the possible $n(n-1)$ edges is included in $\vec{G}$ with probability $p$ independent of any other edges.
\end{definition}

%%% COLOR

\section{Ring Samplers}\label{sec:ring_samp_defs}

% Following \cite{PoPETs:RELSY21}, we define a ring sampler and give an intuition for the measure for the anonymity of rings samplers.

We recall the formal definition of ring samplers~\cite{PoPETs:RELSY21} and define distributions of transaction graphs which are induced by ring samplers.

\begin{definition}[Ring Samplers~\cite{PoPETs:RELSY21}]\label{def:ringsampler}
	A ring sampler $\Samp$ is a (stateless) PPT algorithm which inputs a set of users $U$ and a signer $s \in U$ and outputs a ring $r$ satisfying $s \in r \subseteq U$.
	Syntactically, we write $r \gets \Samp(U, s)$ where $\Samp$ is understood to take uniform randomness which is omitted.
	%If the universe $U$ is clear from the context, we omit $U$ and write $R \gets \Samp(s)$.
\end{definition}

\begin{remark}
	In general, a ring sampler $\Samp$ could input a set $s = \set{s_1,s_2,\ldots} \subseteq U$ of signers and outputs a ring $r$ with $s \subseteq r \subseteq U$.
\end{remark}

Consider the following thought experiment:
Let there be a set of users $U$. At each time $j$, a uniformly random user $s_j$ who has not signed yet decides to issue a ring signature.\footnote{In practice, in case two users publish their signatures simultaneously, a public tie-breaking rule is in place to decide which signature should be verified and accepted first.} To do so, user $s_j$ samples a ring $r_j \gets \Samp(U,s_j)$ and publishes its ring signature together with the ring $r_j$. The ring membership relations of the published rings $r_1, \ldots, r_m$ form a transaction graph, whose distribution is induced by the randomness used for ring sampling.

\begin{definition}[Induced Transaction Graphs]\label{def:ind_tran_graph}
	An induced transaction graph sampler $\cG^\Samp$ is an oracle-aided PPT algorithm which is given access to a ring sampler $\Samp$, inputs a set of users $U$ and a number $m \in [|U|]$ (in unary) of signers, and outputs a transaction graph $G$ and a maximum matching $M$ in $G$.
	The procedures of $\cG^\Samp$ are as described in~\cref{fig:ind_tran_graph}.
	Whenever we are only concerned with the transaction graph $G$ sampled and not the maximum matching $M$, we omit $M$ and write simply $G \gets \cG^\Samp(U,1^m)$.
\end{definition}

In the definition of induced transaction graphs, the maximum matching $M$ output by $\cG^\Samp$ represents the ``true'' signer-ring assignment, i.e. each $(s_j,r_j) \in M$ represents that the ring $r_j$ is sampled by signer $s_j$.

A graph-based deanonymisation attack against (a system employing) a ring sampler $\Samp$ can be modelled by a security experiment involving an adversary $\adv$. We first consider the case with passive adversaries.
The adversary $\adv$ is given a transaction graph $G$, where $(G,M) \gets \cG^\Samp(U,1^{|U|})$ is sampled by an induced transaction graph sampler, %\footnote{Setting $m = |U|$ is without loss of generality, to be explained after~\cref{def:graph-based-anon}.}
and is asked to find an edge $(s^*,r^*)$ of $G$ such that $(s^*,r^*) \in M$, i.e. to correctly identify that the ring $r^*$ is sampled by the signer $s^*$.
For the active setting, we additionally allow the adversary $\adv$ to corrupt a subset $B \subseteq U$ of users.

\begin{figure}
	\begin{pchstack}[boxed,center]

		\procedure[]{$\cG^\Samp(U,1^m)$}
		{
          \\[-1.7em]
          \pcfor j \in [m] \pcdo \\
			\pcind s_j \gets U \setminus \set{s_i}_{i=1}^{j-1} \\
			\pcind r_j \gets \Samp(U,s_j) \\
			R := \tuple{r_1, \ldots, r_m} \\
			E := \set{(u,r) \in U \times R: u \in  r} \\
			M := \set{(s_j, r_j)}_{j \in [m]} \\
			G := (U, R, E) \\
			\pcreturn (G, M)
		}
        \pchspace\pchspace
        		\procedure{$\mathsf{Exp}_{\adv,\Samp}(U)$~\protect\dbox{$\mathsf{Exp}_{\adv,\Samp,\pi}(U)$}\vspace{.275em}}{
			\\[-1.7em]
			\dbox{$B \gets \adv(U)$}\\
			(G = (U,R,E), M) \gets \cG^\Samp(U,|U|) \\
			\begin{subprocedure}
				\dbox{\procedure{}{%
						U \gets U \setminus B\\
						E \gets E \cap (U \times U) \\
						G = (U,R,E)
			}} \end{subprocedure}\\
			(u^*, r^*) \gets \adv(G)\\
			\pcreturn \tuple{(u^*, r^*) \in M}~\dbox{$\land~\tuple{\pi(U,B)  = 1}$}
		}

	\end{pchstack}
	\caption{Induced transaction graph sampler $\cG^\Samp(U,m)$ and Experiments for the security of $\Samp$ against graph-based deanonymisation attacks. The variant incorporating black marble attacks is in dashed boxes.}
	\label{fig:ind_tran_graph}
    \label{fig:exp}
\end{figure}

\begin{definition}\label{def:graph-based-anon}
	Let $\epsilon > 0$ and $\Samp$ be a ring sampler. We say that $\Samp$ is $\epsilon$-secure against graph-based deanonymisation attacks if for any adversary $\adv$ and any set of users $U$,
	\[
		\Pr{\mathsf{Exp}_{\adv,\Samp}(U)} \leq \epsilon.
	\]
	Generalising, for any predicate $\pi$, we say that $\Samp$ is $(\pi,\epsilon)$-secure against graph-based deanonymisation attacks if for any adversary $\adv$ and any set of users $U$,
	\[
		\Pr{\mathsf{Exp}_{\adv,\Samp,\pi}(U)} \leq \epsilon.
	\]
	where the experiments $\mathsf{Exp}_{\adv,\Samp}$ and $\mathsf{Exp}_{\adv,\Samp,\pi}$ are described in~\cref{fig:exp}.
\end{definition}

In~\cref{def:graph-based-anon}, we assume that all users have signed, i.e. $m = |U|$. This captures worst case security since the security experiment for smaller $m$ can be emulated by the worst case adversary, as we will show in \cref{thm:balance_is_upper_bound}. While $\mathsf{Exp}_{\adv,\Samp}$ captures passive attacks, $\mathsf{Exp}_{\adv,\Samp,\pi}$ further captures active attacks by allowing the adversary to corrupt a subset $B$ of users prior to receiving the transaction graph with the restriction that $(U,B)$ satisfies the predicate $\pi$. Setting $\pi$ to only accept $B = \emptyset$, we recover the passive case.

Note that a trivial strategy for graph-based deanonymisation is to pick $r^*$ with the fewest members, pick a random member $s^* \gets r^*$, and output $(s^*,r^*)$. Clearly, this strategy has success probability $1/|r^*| = 1/(\min_{r \in R} |r|)$.
As we will show in \cref{sec:interpretation}, conditioned on $G = \core{G}$, this is in fact the best strategy for attacking against partitioning ring samplers.

%% file: fig_induced_digraph.tex
% !TEX root = main.tex

\begin{figure}[h]
    \centering
    \begin{tikzpicture}[every node/.style={circle,draw,minimum size=.3cm,inner sep=0pt}]
    \foreach \i in {1,...,8}{
    	\node[label=left:$u_{\i}$] (s\i) at (0,-\i) {};
    }
    \foreach \i in {1,...,7}{
    	\node[label=right:$r_{\i}$] (r\i) at (2,-\i) {};
    	\draw[ultra thick,yellow!80!gray!60] (s\i) -- (r\i);
    }
    \begin{scope}[ultra thick,color=black]
    	\draw (s2) -- (r4);
    	\draw (s3) -- (r4);
    	\draw (s4) -- (r5);
    	\draw (s4) -- (r6);
    \end{scope}
    \begin{scope}[ultra thick,color=cyan]
    	\draw (s1) -- (r2);
    	\draw (s2) -- (r1);
    	\draw (s2) -- (r3);
    	\draw (s3) -- (r1);
    	\draw (s6) -- (r7);
    	\draw (s7) -- (r6);
    \end{scope}
    \begin{scope}[ultra thick,color=red!80]
    	\draw (s6) -- (r5);
    	\draw (s8) -- (r7);
    	\draw[dashed] (s6) -- (r7);
    	\draw[dashed] (s7) -- (r6);
    \end{scope}
    \draw[thick,dotted] (-.8,-.6) rectangle (2.8,-7.4);
    %%%%%%%%%%%%%%%%%%%%%%%%%%%%%%%%%%
    \begin{scope}[shift={(4,.3)},yscale=1.35,>=stealth']
    	\node[label=above:$1$] (n1) at (1,-1) {};
    	\node[label=left:$2$] (n2) at (0,-2) {};
    	\node[label=right:$3$] (n3) at (2,-2) {};
    	\node[label=right:$4$] (n4) at (1,-3) {};
    	\node[label=left:$5$] (n5) at (0,-4) {};
    	\node[label=right:$6$] (n6) at (2,-4) {};
    	\node[label=left:$7$] (n7) at (1,-5) {};
    	\node[label=below:$8$] (n8) at (1,-6) {};
    	\begin{scope}[->,ultra thick,color=black]
    		\draw (n2) -- (n4);
    		\draw (n3) -- (n4);
    		\draw (n4) -- (n5);
    		\draw (n4) -- (n6);
    	\end{scope}
    	\begin{scope}[->,ultra thick,color=cyan]
    		\draw (n1) edge[bend right] (n2);
    		\draw (n2) edge[bend right] (n1);
    		\draw (n2) edge[bend right] (n3);
    		\draw (n3) edge[bend right] (n1);
    		\draw (n6) edge[bend right] (n7);
    		\draw (n7) edge[bend right] (n6);
    	\end{scope}
    	\begin{scope}[->,ultra thick,color=red]
    		\draw (n6) -- (n5);
    		\draw (n8) -- (n7);
    		\draw[dashed,-{Stealth[round,length=9pt]}] (n6) edge[bend right] (n7);
    		\draw[dashed,-{Stealth[round,length=9pt]}] (n7) edge[bend right] (n6);
    	\end{scope}
    \end{scope}
    \end{tikzpicture}
    \caption{Example of a transaction graph $G$ ($U$ and $R$ being nodes on left and right respectively) and its induced digraph $\ID{G}$. The subgraph in the dotted rectangle is $G^\upp$. The yellow, blue and red edges correspond to edges considered in \cref{lem:tassa} \cref{item:ii} to \cref{item:lownode} respectively, the black edges are none of them.}
    \label{fig:induced_digraph}
\end{figure}
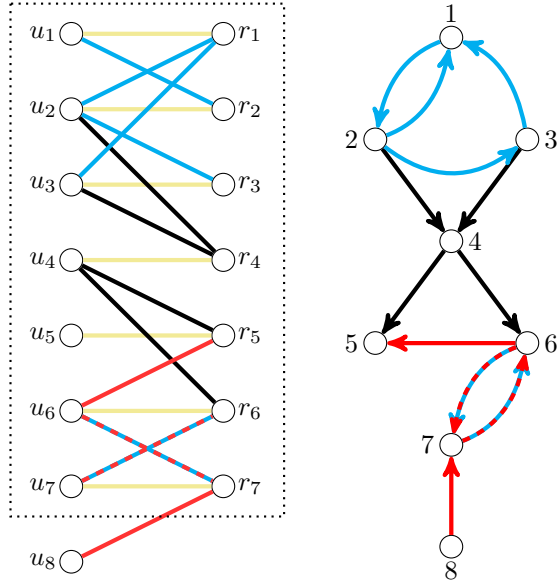

%% file: connectivity.tex
% !TEX root = main.tex

\section{From Cores to Induced Digraphs}\label{sec:reduction}

In this section we reduce the problem of upper-bounding $\Pr{\rvG\neq\core{\rvG}}$ to a problem concerning the strong connectivity of digraphs. We first recall a result from Tassa~\cite{TCS:Tassa12} for general bipartite graphs specialised to the case of transaction graphs.

\begin{lemma}[Tassa~\cite{TCS:Tassa12}]\label{lem:tassa}
	Let $G=(U,R,E)$ be a transaction graph, where $U = \set{u_i}_{i=1}^n$ and $R = \set{r_j}_{j=1}^m$, and $M = \set{(u_j, r_j)}_{j=1}^{m}$ be a maximum matching in $G$.
	The core $\core{G} = (U,R,E')$ is a transaction graph where $E'$ is the union of the following sets:
	\begin{enumerate}
		\item \label{item:ii} The maximum matching $M$,
		\item \label{item:scc} $\set{(u_i,r_j): (i,j)~\text{is in some SCC of}~\id_M(G)}$, and
		\item \label{item:lownode} $\set{(u_i,r_j): \begin{aligned}
			&(i,j)~\text{is reachable from} \\[-4pt]
			&\id_M(G) \setminus \id_M(G^M) ~\text{through}~\id_M(G)
	\end{aligned}}$.
	\end{enumerate}
\end{lemma}

\begin{proof}
	This is a direct summary of the results in Tassa~\cite{TCS:Tassa12}, specifically Theorem 2.2 and Algorithm 2 for \cref{item:scc}, and Proposition 2.4, Theorem 2.7 and Algorithm 3 for \cref{item:lownode}. \cref{item:ii} is obvious by definition.
	% For imbalanced transaction graph, using the terminologies from the paper, set $R$ as left nodes and $S$ as right nodes, and there is no left-to-right digraph to be considered.
\end{proof}

In the example given in \cref{fig:induced_digraph}, the edges considered in \cref{lem:tassa} \cref{item:ii,item:scc,item:lownode} are coloured yellow, blue, and red respectively. The black edges are those not in $\core{G}$, corresponding to impossible signer-signature assignments that can be ruled out.

By \cref{lem:tassa} \cref{item:lownode}, any edge $(u_i,r_j)$ of an imbalanced transaction graph with $i > m$ is maximum-matchable.
Further, note that an edge in a digraph must either be within an SCC or connecting two SCCs, and not both.
%Hence, any edge $(i,j)$ being not in \cref{lem:tassa} \cref{item:scc} implies $(i,j)$ is an edge connecting two SCCs in $\vec{G}$.
Hence, from \cref{lem:tassa} \cref{item:scc} and \cref{item:lownode}, any edge $(u_i,r_j)$ not being in $\core{G}$ implies $(u_i,r_j)$ is an edge connecting two SCCs in $\vec{G}^M$.

%\begin{lemma}\label{lem:GeqCore}
%	For any transaction graph $G$, if $\core{G}$ is connected then $G=\core{G}$.
%\end{lemma}
%
%\begin{proof}
%	Suppose $G\neq\core{G}$, so there is an edge $(i,j)$ in $G$ being not in $\core{G}$. From \cref{lem:tassa} $(i,j)$ must be an edge connecting two SCCs in $\vec{G}$. Call the sets of nodes in the two SCCs $C_1$ and $C_2$. Since any edge $(i,j)$ where $i \in C_1$ and $j \in C_2$ are not in $\core{G}$, the sets of nodes $\set{i, r_i}_{i \in S_1}$ and $\set{i, r_i}_{i \in S_2}$ are disconnected in $\core{G}$.
%\end{proof}

%Denote by $\rvG=(S,R,\rvE)$ the random transaction graph induced by a ring sampler, where $\rvE$ is the random existence of edges in $\rvG$, and $\rvD$ be the distribution which $\rvG$ follows. %, and $\rvC$ be the random variable denoting $\rvG$'s set of connected components.

Using \cref{lem:tassa}, we derive in the following a number of lemmas on the probability of $G\neq\core{G}$, which together will lead to \cref{thm:core}.

We begin with~\cref{lem:prodH}, which states that if $G$ has a partition $P$, then the cores $\core{H}$ of the chunks $H \in P$ collectively tell us everything about $\core{G}$.

\begin{lemma}\label{lem:prodH}
	Let $G$ be a transaction graph and $P$ be a partition of $G$.
	It holds that $G=\core{G}$ if and only if $H = \core{H}$ for all $H \in P$.
\end{lemma}

\begin{proof}
	Recall that $G = \bigcup_{H \in P} H$. Suppose for the moment that $\core{G} = \bigcup_{H \in P} \core{H}$, then we can prove the lemma statement as follows.

	Suppose $G = \core{G}$.
	We have $\bigcup_{H \in P} H = \bigcup_{H \in P} \core{H}$.
	Observe that for distinct $H, H' \in P$ we must have $H \cap H' = \emptyset$.
	Therefore $H = \core{H}$ for all $H \in P$.

	Suppose $H = \core{H}$ for all $H \in P$, then $G = \bigcup_{H \in P} H = \bigcup_{H \in P} \core{H} = \core{G}$.

	It remains to show that $\core{G} = \bigcup_{H \in P} \core{H}$.

	Let the edge $e \in \core{G}$, \ie $e$ belongs to a maximum matching $M$ in $G$.
	Since $\core{G} \subseteq G$, and $P$ is a partition of $G$, we have $e \in H^*$ for some $H^* \in P$.
	Since $M \cap H^*$ is a maximum matching in $H^*$, we have $e \in \core{H^*}$.
	This shows that $\core{G} \subseteq \bigcup_{H \in P} \core{H}$.

	Let the edge $e \in \core{H}$ for some $H \in P$, \ie $e$ belongs to a maximum matching $Y$ in $H$. Let $M$ be a maximum matching in $G$ whose existence is guaranteed since $G$ is a transaction graph. Then $M^* := (M \setminus H) \cup Y$ is also a maximum matching in $G$.
	Consequently $e \in M^* \subseteq \core{G}$, which implies $\bigcup_{H \in P} \core{H} \subseteq \core{G}$.
\end{proof}

As an immediate corollary of~\cref{lem:prodH}, \cref{cor:prodH} states a similar relation concerning distributions of transaction graphs. In particular, it states that the probability of $G \neq \core{G}$ is upper-bounded by the probability of the existence of a chunk $G_C$ of $G$ with $G_C \neq \core{G_C}$, which can further be upper-bounded by the union bound.

\begin{corollary} \label{cor:prodH}
	Let $\mathcal{G}$ be any distribution of transaction graphs with identical vertex sets and let $\set{\mathcal{G}_C}_{C \in P}$ be a partition of $\mathcal{G}$.
	% Consider a transaction graph $G$ which follows distribution .
	% Suppose that there exists a partition $P$ of $G$ in the support of $\mathcal{G}$, such that for all $\rvH \in P$, $\rvH$ is a transaction graph following some distribution $\mathcal{H}$. %and that $\rvH$'s are pairwise independent of each other.
	%Assume that in the support of $\rvD$, there exists a partition $\rvP = \{v_1,\ldots,v_\ell\}$ of the set of nodes $S \cup R$ for some $\ell \in \NN$, such that
	%$v_i$ and $v_j$ are disconnected in $\rvG$ for all distinct $v_i,v_j \in \rvP$.
	%Let $\rvC$ be the set of subgraphs of $\rvG$, where $\rvH = (v, v \times v \cap \rvE)$ for all $\rvH \in \rvC$, $v \in \rvP$, so that $\cup_{\rvH \in \rvC} \rvH = \rvG$.
	%$\mathcal{G}_i = (v_i, v_i \times v_i \cap \rvE)$ for all $v_i \in \rvP$ and $\cup_{v_i \in \rvP} \mathcal{G}_i = \rvG$. Let $\rvC := \{\mathcal{G}_i\}_{v_i \in \rvP}$.
	Then
	\begin{align*}
	\PrG{\rvG \neq \core{\rvG}}
	\leq \sum_{C \in P} \Pr[G_C \gets \mathcal{G}_C]{G_C \neq \core{G_C}}.
	\end{align*}
\end{corollary}

\begin{proof}
	By \cref{lem:prodH}, we have
	\begin{align*}
	\PrG{\rvG \neq \core{\rvG}}
	= \PrG{\exists C \in P,~G_C \neq \core{G_C}}
	\end{align*}
	where on the right hand side $\set{G_C}_{C \in P}$ is a partition of $G$.
	We then arrive at the desired conclusion by applying the union bound.
\end{proof}

Next, \cref{lem:boundGu} upper-bounds the probability of $G \neq \core{G}$ by that of $G^\upp \neq \core{G^\upp}$, where we recall that $G^\upp$ is an arbitrary fixed upper graph of $G$. Note that $G^\upp$ is balanced by definition. Therefore,~\cref{lem:boundGu} in some sense means that balanced transaction graphs are the worst cases for how likely transaction graphs are equal to their respective cores.

\begin{lemma}\label{lem:boundGu}
	Let $G=(U,R,E)$ be a transaction graph.
	If $G^\upp = \core{G^\upp}$, then $G = \core{G}$.
	Consequently, let $\mathcal{G}$ be any distribution of transaction graphs, we have
	\[
		\PrG{\rvG \neq \core{\rvG}} \leq \PrG{\rvG^\upp \neq \core{\rvG^\upp}}.
	\]
\end{lemma}

\begin{proof}
%	For balanced graph the statement is trivial.
%	For imbalanced graph consider the following.
%	Let $\vec{G}$ and $\vec{G}'$ be the induced digraphs of $G$ and $G'$ respectively.
	Let $M$ be a maximum matching in $G$ such that $G^\upp = G^M = (U^M, R, E^M)$.
	It suffices to show that each chunk in the partition $\set{E \setminus E^M, E^M \setminus M, M}$ of $E$ is a subset of the edges in $\core{G}$.

	First, we have $e\in\core{G}$ for all $e \in M$ by the definition of core.
	Moreover, by \cref{lem:tassa} \cref{item:lownode}, $e\in\core{G}$ for edge $e \in E \setminus E^M$.

	It remains to consider $E^M \setminus M$.
	Given that $G^M = \core{G^M}$, all $e \in E^M$ are in $\core{G^M}$.
	Since $G^M$ is balanced, from \cref{lem:tassa} we have that all $e\in E^M \setminus M$ are in some SCC of $\id_M(G^M)$.
	By construction, an SCC in $\id_M(G^M)$ is also an SCC in $\id_M(G)$, so by \cref{lem:tassa} \cref{item:scc} all $e \in E^M \setminus M$ are also in $\core{G}$.
	% We conclude that $e\in\core{G}$ for all $e \in G$, therefore $G = \core{G}$.
%
%	Next suppose $\core{G'}$ is not connected, say some edge $(i,j) \in E$ is not in $\core{G'}$. Let $\vec{G}$ and $\vec{G'}$ be $G$ and $G'$'s induced digraphs respectively. By \cref{lem:tassa} $(i,j)$ must be an edge connecting two SCCs in $\vec{G'}$, hence the same in $\vec{G}$. Being not in \cref{lem:tassa} \cref{item:scc}, $(i,j)$ could however still be in $\core{G}$ by being in \cref{item:lownode}. When all edges not in \cref{item:scc} are however in \cref{item:lownode} so that $G=\core{G}$, $\core{G}$ is connected by \cref{lem:GeqCore}.
\end{proof}

Our last lemma for this section,~\cref{lem:sc}, upper-bounds the probability of $G \neq \core{G}$ by that of $\id(G)$ being not strongly connected, where we recall that $\id(G)$ is an induced digraph of $G$ with arbitrarily chosen maximum matching.

\begin{lemma}\label{lem:sc}
	%	For a balanced and connected transaction graph $G$, $\core{G}$ is connected if and only if its induced digraph $\vec{G}$ is strongly connected.
	Let $G$ be a transaction graph.
	If $\id(G)$ is strongly connected, then $G=\core{G}$.
	Furthermore, if $G$ is both balanced and connected, then the converse also holds.
	Consequently, let $\mathcal{G}$ be any distribution of transaction graphs, we have
	\[
		\PrG{\rvG \neq \core{\rvG}} \leq \PrG{\id(\rvG) \notin \SC},
	\]
	and the inequality become equality if $\mathcal{G}$ is a distribution of balanced and connected transaction graphs.
\end{lemma}

\begin{proof}
	If $\id(G)$ is strongly connected, then by \cref{lem:tassa} all edges in $G$ are in $\core{G}$, hence $G=\core{G}$.

	The second statement is proven by contraposition. Suppose $\id(G)$ is not strongly connected, so it has at least two SCCs $\vec{C}_1$ and $\vec{C}_2$.
	If $G$ is connected, then $\id(G)$ is by construction weakly connected, and there exists an edge $(i,j)$ in $\id(G)$, where $i$ is a node of $\vec{C}_1$ and $j$ is a node of $\vec{C}_2$. By \cref{lem:tassa} we have that $(u_i,r_j)$, which is an edge in $G$, is not in $\core{G}$, hence $G \neq \core{G}$.
	%	From \cref{lem:tassa}, any edge $(i,j)$ in $\vec{G}$ where $i \in C_1$ and $j \in C_2$ implies $(i,j)$ is not in $\core{G}$.
	%	We conclude there is no edge $(i,j)$ in $\core{G}$ where $i \in C_1$ and $j \in C_2$, so the sets of nodes $\set{i, r_i}_{i \in C_1}$ and $\set{i, r_i}_{i \in C_2}$ are disconnected in $\core{G}$. %$\core{G}$ must not be connected.
\end{proof}

Note that by construction, $\id(G)$ is strongly connected only if $G$ is balanced.
Therefore the inequality in \cref{lem:sc} becomes trivial if $G$ is imbalanced.

Chaining together the above lemmas, we arrive at the main theorem of this section, which upper-bounds the probability of $G \neq \core{G}$ by a sum of probabilities related to the strong connectivity of the induced digraphs of the chunks of $G$.

\begin{theorem}\label{thm:core}
	Let $\mathcal{G}$ be any distribution of transaction graphs and let $\set{\mathcal{G}_C}_{C \in P}$ be a partition of $\mathcal{G}$.
	Then
	\begin{align*}
		\PrG{\rvG \neq \core{\rvG}}
		\leq \sum_{C \in P} \Pr[G_C \gets \mathcal{G}_C]{\id(G^\upp_C) \notin \SC}.
	\end{align*}
%	where $\vec{G}^M_i$ is the induced digraph of $G^M_i$.
\end{theorem}

\begin{proof}
	From \cref{cor:prodH},
	\[
		\PrG{\rvG \neq \core{\rvG}}
		%= 1 - \prod_{\rvH \in P} \Pr{\rvH = \core{\rvH}}.
		\leq \sum_{C \in P} \Pr[G_C \gets \mathcal{G}_C]{G_C \neq \core{G_C}}.
	\]
	From \cref{lem:boundGu,lem:sc} we have
	\begin{align*}
	\Pr[G_C \gets \mathcal{G}_C]{G_C \neq \core{G_C}}
	\leq &\Pr[G_C \gets \mathcal{G}_C]{G^\upp_C \neq \core{G^\upp_C}} \\
	\leq &\Pr[G_C \gets \mathcal{G}_C]{\id(G^\upp_C) \notin \SC}
	\end{align*}
	for any $C \in P$.
	Combining the above yields the desired result.
\end{proof}

%% file: partitioning.tex
% !TEX root = main.tex

\section{Induced Transaction Graphs}\label{sec:partitioning_samplers}

Our goal in this section is to obtain a candidate upper bound for $\Pr{G \neq \core{G}}$, where $G$ is a random transaction graph induced by a (regular) partitioning sampler~\cite{PoPETs:RELSY21}.
%For this, we first formalise the notion of ring-sampler-induced transaction graphs.
For this, we first prove a theorem on the sufficiency of considering balanced induced transaction graphs.
We then recall the definition of (regular) partitioning samplers~\cite{PoPETs:RELSY21} and apply the established theorems.
%, and propose a variant called binomial partitioning samplers.
We realise that $\Pr{G \neq \core{G}}$ can be upper-bounded in terms of $\Pr{\vec{G} \notin \SC}$ where $\vec{G}$ is sampled from $\vec{\cG}^{\mathtt{reg}}_{k,n}$ (recall \cref{def:regD}).

%Using the results in~\cref{sec:reduction}, to upper-bound $\Pr{G \neq \core{G}}$, it suffices to upper-bound $\Pr[\vec{G} \gets \cG^{\mathtt{reg}}_{k,n}]{\vec{G} \notin \SC}$. For the latter, we conjecture that
%\begin{align*}
%	\rho^{\mathtt{reg}}_{k,n} := \Pr[\vec{G} \gets \vec{\cG}^{\mathtt{reg}}_{k,n}]{\vec{G} \notin \SC}
%	\leq
%	\Pr[\vec{G} \gets \vec{\cG}^{\mathtt{bin}}_{p,n}]{\vec{G} \notin \SC} =: \rho^{\mathtt{bin}}_{p,n}
%\end{align*}
%for $p = \frac{k}{n-1}$.
%Following through, we realise that
%\[\lim_{n \to \infty} \rho^{\mathtt{bin}}_{p(n),n}\]
%is known for certain choices of $p(n)$.
%Based on the expression of the limit, we heuristically obtain a candidate upper bound of $\rho^{\mathtt{bin}}_{p,n}$.
%
%Although we are not able to prove the above conjectures analytically, in~\cref{sec:exp} we show empirical evidence that they seem to hold for all choices of $k$ and $n$ we considered.

\subsection{Balanced Transaction Graphs}

%We define the notion of ring-sampler-induced transaction graphs. Specifically, for some ring sampler $\Samp$ and $M \subseteq N \subseteq U$, we define the distribution $\cG^\Samp_{N,M}$ of transaction graphs induced by $\Samp$ and $(N,M)$. %In case $(N,M) =(U,U)$, we write $\cG^\Samp$ for $\cG^\Samp_{U,U}$.

%\begin{definition}[Induced Transaction Graphs]\label{def:ind_tran_graph}
%	For $M \subseteq N \subseteq U$, the distribution $\cG^{\Samp}_{N,M}$ of transaction graphs induced by a ring sampler $\Samp$ is
%	\begin{align*}
%		\cG^{\Samp}_{N,M}  :=
%		\set{\tuple{N,M,\set{(i,j) \in N \times M: i \in \Samp(U,j)}}}.
%	\end{align*}
%	If $(N,M) = (U,U)$, we simply write $\cG^{\Samp}$ for $\cG^{\Samp}_{U,U}$.
%\end{definition}
%
%Note that $\cG^{\Samp}_{N,M}$ is a distribution with randomness induced from that of $\Samp$, and all transaction graphs in the support of $\cG^\Samp$ are balanced.

Intuitively, it is easier for an adversary to deanonymise signers when more information about them is available, for example, when more rings sampled by the signers are given. Following this line of thought, an adversary should be successful in  deanonymising signers with the highest probability when all users have signed.
%For example, it should hold that
%$\Pr[G \gets \cG^\Samp_{N,M}]{G \neq \core{G}}
%	\leq \Pr[H \gets \cG^\Samp]{H \neq \core{H}}$.

To formalise this claim, we first prove a technical lemma which states that, if $H$ is constructed by adding ring nodes to a transaction graph $G$, then $G \neq \core{G}$ implies $H \neq \core{H}$. %The claim then follows straightforwardly from the technical lemma.

\begin{lemma} \label{lem:ringsize}
	Let $G = (U,R,E)$ and $H = (U, R', E')$ be transaction graphs where $R \subset R'$ and $E = E' \cap (U \times R)$, \ie $H$ can be constructed from $G$ by adding ring nodes $R' \setminus R$ and edges connecting the new ring nodes to some signer nodes $U$. If $G \neq \core{G}$, then $H \neq \core{H}$.
\end{lemma}

\begin{proof}
	%For the ease of notation, let $n = |N|$, $m = |M|$ and $m' = |M'|$, so that $m < m' \leq n$.
	Let $U = \set{u_i}_{i=1}^n$,
	$R = \set{r_j}_{j=1}^m$,
	and $M = \set{(u_j,r_j)}_{j=1}^m$ be a maximum matching in $G$.
	It suffices to prove the case $|R'| = |R|+1$, and the lemma follows by induction.
	We therefore assume from here on $R' = R \cup \set{r_{m+1}}$ where $r_{m+1} \notin R$.

	Let $M' := M \cup \set{(u_{m+1},r_{m+1})}$ be a maximum matching in $H$.
	Let $\ID[M]{G} = ([n], F)$ and $\ID[M']{H} = ([n], F')$. %be the induced digraphs of $G$ and $H$ respectively, and $\vec{G}^\upp$ and $\vec{H}^\upp$ be the induced digraphs of $G^\upp$ and $H^\upp$ respectively.
	Note that $F \subseteq F'$ (and hence $\ID[M]{G} \subseteq \ID[M']{H}$), with the new edges in $F' \setminus F$ being of the form $(i, m+1)$ where $i \in [n] \setminus \set{m+1}$.
	%Note also that $\vec{G}\setminus \vec{G^\upp} = \tuple{\set{m',m'+1,\ldots,n}, \emptyset}$ and $\vec{H} \setminus \vec{H}^\upp = \tuple{\set{m'+1,\ldots,n}, \emptyset}$, therefore $\tuple{\vec{H} \setminus \vec{H}^\upp} \subseteq \tuple{\vec{G}\setminus \vec{G}^\upp}$.

	Suppose $G \neq \core{G}$, so there exists an edge $e^* = (u_{i^*}, r_{j^*})$ in $G$ which is not in $\core{G}$.
	From \cref{lem:tassa} we have $i^* \neq j^*$, therefore $e^* \in \ID[M]{G} \subseteq \ID[M']{H}$.
	We prove in the following that $e^*$ is not in any SCC of $\ID[M']{H}$, and $e^*$ is not reachable from $\ID[M']{H} \setminus \ID[M']{H^{M'}}$ through $\ID[M']{H}$.
	Hence, by \cref{lem:tassa}, $e^*$ is not in $\core{H}$, and $H \neq \core{H}$.

	%From~\cref{lem:tassa}, we know that $i^\ast \neq j^\ast$, and the edge $(i^\ast, j^\ast)$ is an edge in $\vec{G}$ which connects two SCCs of $\vec{G}$. Call the sets of nodes in the two SCCs $\vec{C}_1$ and $\vec{C}_2$, so $i^\ast \in \vec{C}_1$ and $j^\ast \in \vec{C}_2$. Note that $\vec{C}_1, \vec{C}_2 \in [m]$.

	%we know that $i \neq j$, $(i,j)$ does not belong to any SCC of $\vec{G}$, and $(i,j)$ is not reachable from $\vec{G}\setminus\vec{G}^\upp$ through $\vec{G}$.
	%In other words, $(i, j)$ is connecting $\vec{C}_1$ to $\vec{C}_2$, where $\vec{C}_1$ and $\vec{C}_2$ are distinct SCCs of $\vec{G}$ whose vertex sets are subsets of $[m]$.
	%This in particular implies that $i,j \in [m]$.

	%We claim that $(s_{i^\ast}, r_{j^\ast})$ does not belong to $\core{H}$, so that $H \neq \core{H}$. To prove this claim, according to~\cref{lem:tassa}, it suffices to show that

	We first show that $e^*$ is not in any SCC of $\ID[M']{H}$.
	For this, note that from~\cref{lem:tassa}, $e^*$ is an edge which connects two SCCs of $\ID[M]{G^M}$.
	Let $\vec{C}$ be an SCC of $\ID[M]{G^M}$ such that $i^*$ is a node of $\vec{C}$.
	Observe that by construction, $\vec{C}$ is also an SCC of $\ID[M]{G}$.
	Now, since the vertex set of $\vec{C}$ is subset of $[m]$ (the vertex set of $\ID[M]{G^M}$), there is no edge $(i,j) \in F'\setminus F$ with node $j$ in $\vec{C}$ (since all edges in $F' \setminus F$ are of the form $(i,m+1)$).
	Clearly this implies, first, that there is no edge in $F'\setminus F$ with both ends in $\vec{C}$, and second, that there is no edge in $F'\setminus F$ which connects from any node $v \in \ID[M']{H} \setminus \vec{C}$ to $\vec{C}$.
	Therefore, $\vec{C}$ remains an SCC in $\ID[M']{H}$ by definition, and it follows that $e^*$ is not in any SCC of $\ID[M']{H}$.

%	Recall that $(i, j)$ is connecting $\vec{C}_1$ to $\vec{C}_2$, where $\vec{C}_1$ and $\vec{C}_2$ are distinct SCCs of $\vec{G}$.
%	It suffices to show that $\vec{C}_1$ remains an SCC of $\vec{H}$.
%	For this, it suffices to show that $\vec{C}_1$ remains an SCC of $\vec{H}$.
%	Suppose not, let $\vec{C}'_1$ be an SCC of $\vec{H}$ with $\vec{C}_1 \subset \vec{C}'_1$.
%	There must exist two edges $(i, m')$ and $(m', j)$ in $\vec{C}'_1$ such that $i, j \in [m]$ are nodes in $\vec{C}_1$.
%	The existence of the edge $(m', j)$, however, contradicts the observation that all edges which are added to $\vec{G}$ to form $\vec{H}$ must be of the form $(i,m')$.
%

	We next show that $e^*$ is not reachable from $\ID[M']{H} \setminus \ID[M']{H^{M'}}$ through $\ID[M']{H}$. We begin by drawing attention to two points.
	First, by \cref{lem:tassa}, $e^*$ is not reachable from $\ID[M]{G}\setminus \ID[M]{G^M}$ through $\ID[M]{G}$. Second, $e^*$ is not reachable from node $m+1$ through $\ID[M']{H}$, since $e^*$ is not reachable from $m+1$ through $\ID[M]{G}$ and all edges in $F' \setminus F$ are of the form $(i,m+1)$.
	We proceed to prove the statement by contradiction. Suppose $e^*$ is reachable from $\ID[M']{H} \setminus \ID[M']{H^{M'}}$ through $\ID[M']{H}$, then there exists a directed path $P = \set{(v_{i-1},v_{i})}_{i=1}^\ell$ in $\ID[M']{H}$, %from $\vec{H} \setminus \vec{H}^\upp$ to node $i^*$ through $\vec{H}$,
	where $v_0$ is a node of $\ID[M']{H} \setminus \ID[M']{H^{M'}}$, $(v^{\ell-1},v^\ell) = e^*$, and node $v_i \in R'$ for all $i\in [\ell]$.\footnote{The condition on the intermediate nodes can be achieved by first considering any path $P$ from node $v_0 \in \ID[M']{H} \setminus \ID[M']{H^{M'}}$ to $e^*$ through $\ID[M']{H}$, and then taking the tail of $P$ such that no intermediate node in the tail belongs to $\ID[M']{H} \setminus \ID[M']{H^{M'}}$.}
	However, $v_i \neq m+1$ for all $i \in [\ell]$, otherwise contradicting that $e^*$ is not reachable from node $m+1$ through $\ID[M']{H}$. Therefore $v_i \in M$ for all $i\in [\ell]$.
	Since all edges in $F' \setminus F$ are of the form $(i,m+1)$, we now have that all edges in $P$ belong to $\ID[M]{G}$.
	In other words, $e^*$ is reachable from $\ID[M']{H} \setminus \ID[M']{H^{M'}}$ through $\ID[M]{G}$.
	Finally, since $\ID[M']{H} \setminus \ID[M']{H^{M'}} = \tuple{[n] \setminus (R \cup \set{m+1}), \emptyset} \subset \tuple{[n] \setminus R, \emptyset} = \ID[M]{G}\setminus \ID[M]{G^M}$, we arrive at that $e^*$ is reachable from $\ID[M]{G}\setminus \ID[M]{G^M}$ through $\ID[M]{G}$, a contradiction.
\end{proof}

From \cref{lem:ringsize} we obtain our next theorem, which states that for any number of signers $m \leq |U|$, $\Pr{G \neq \core{G}}$ is upper-bounded by that when $m =|U|$, \ie the case that all users have signed.

\begin{theorem}\label{thm:balance_is_upper_bound}
	For any ring sampler $\Samp$ and any $m \leq |U|$, it holds that
	\begin{align*}
		&\Pr[G \gets \cG^\Samp(U,1^m)]{G \neq \core{G}} \\
		\leq& \Pr[H \gets \cG^\Samp(U,1^{|U|})]{H \neq \core{H}}.
	\end{align*}
\end{theorem}

\begin{proof}
	As $\Samp$ is stateless, the distributions of the outputs of independent runs of $\Samp$ are independent.
	Hence, referring to \cref{fig:ind_tran_graph}, sampling $H$ from $\cG^\Samp(U,1^{|U|})$ is equivalent to first running the for-loop in $\cG^\Samp(U,1^{|U|})$ only up to $j=m$ to sample $G$, then running the remaining of the loop to sample $G'$, and outputting $H := G \cup G'$.
	From~\cref{lem:ringsize}, we know that $H \neq \core{H}$ whenever $G \neq \core{G}$. The claim thus follows immediately.
\end{proof}

\subsection{Regular Partitioning Samplers} \label{sec:sampreg}

We consider a special case of the partitioning samplers defined in~\cite{PoPETs:RELSY21}, where there is only one public partition of $U$ and only one signer per ring. The general case with a distribution of partitions and more than one signer can be handled with generic techniques~\cite{PoPETs:RELSY21}.
Such partitioning samplers, which we refer to as the regular partitioning samplers $\RegSamp[P,k]$, are parametrised by the partition $P$ of $U$ and a number of decoys $k \in \NN$ for each ring, such that $k < |C|$ for each chunk $C \in P$. We recall its definition below.

\begin{quote}
	$\RegSamp[P, k](U,s)$:
	Initiate $r := \set{s}$.
	Let $C \in P$ be the unique chunk containing $s$.
	Sample a uniformly random $k$-subset $r' \subseteq C \setminus \set{s}$.
    Output $r := r \cup r'$.
\end{quote}

We observe that a $\RegSamp[P,k]$-induced transaction graph $G$ takes a special form -- it can be partitioned into independent subgraphs $\set{G_C}_{C \in P}$, each representing the induced transaction graph of a chunk in $P$.
Moreover, if a subgraph $G_C$ is balanced, then its induced digraph $\ID{G_C}$ is a $k$-in-degree regular digraph.
% Since a regular partitioning sampler adopts uniform sampling, we arrive at the following proposition relating $\cG^{\RegSamp[P,k]}$ and $\vec{\cG}^\regtag_{k,|C|}$ for $C \in P$.
We therefore arrive immediately at the following lemma.

% \begin{proposition} \label{pro:Greg}
% %	Let $P$ be such that $|C| = n$ for each $C \in P$.
% 	For the partition $\set{\cG^{\RegSamp[P,k]}_{C,C}}_{C \in P}$ of $\cG^{\RegSamp[P, k]}$,
% 	it holds that $\ID{\cG^{\RegSamp[P,k]}_{C,C}} = \vec{\cG}^\regtag_{k,|C|}$ for each $C \in P$.
% \end{proposition}

%In the above, $\ID{\cG^{\RegSamp[P,k]}_{C,C}} = \vec{\cG}^\regtag_{k,|C|}$ means that sampling $\vec{G}$ from $\vec{\cG}^\regtag_{k,|C|}$ is equivalent to first sampling $G$ from $\cG^{\RegSamp[P,k]}_{C,C}$ and then computing $\vec{G} = \ID{G}$.

\begin{lemma}\label{pro:Greg}\label{cor:Greg}
	Let $U$ be a set of users and $P$ be a partition of $U$.
	Let $k \in \NN$ such that $k < |C|$ for each $C \in P$.
	Write $\Samp := \RegSamp[P, k]$.
	For any $m \leq |U|$, %it holds that
	\begin{align*}
	\Pr[G \gets \cG^{\Samp}(U,1^m)]{G \neq \core{G}}
	\leq \sum_{C \in P} \Pr[\vec{G} \gets \vec{\cG}^\regtag_{k,|C|}]{\vec{G} \notin \SC}.
	\end{align*}
\end{lemma}

\begin{proof}
	\qquad $\begin{aligned}[t]
	\Pr[G \gets \cG^{\Samp} (U,1^m)]{G \neq \core{G}} 
	\leq& \Pr[G \gets \cG^{\Samp} (U,1^{|U|})]{G \neq \core{G}} \\
	\leq& \sum_{C \in P} \Pr[G \gets \cG^{\RegSamp[\set{C},k]} (C,1^{|C|})]{\ID{G} \notin \SC} \\
	=& \sum_{C \in P} \Pr[\vec{G} \gets \vec{\cG}^\regtag_{k,|C|}]{\vec{G} \notin \SC},
	\end{aligned}$\\
	where the first inequality follows from \cref{thm:balance_is_upper_bound}, the second inequality from \cref{thm:core}, and the equality follows from direct inspection.
\end{proof}

\Cref{pro:Greg} relates the probability of $G \neq \core{G}$ with that of $\vec{G} \notin \SC$, where $G$ is a transaction graph induced by a regular partitioning sampler and $\vec{G}$ is a $k$-in-degree regular digraph.
Unfortunately, the strong connectivity of random $k$-in-degree regular digraphs seems to be a non-trivial problem~\cite[Problem 38]{mauldin1981scottish}. While (asymptotic) results on the connectivity of random $k$-regular (undirected) graphs are established~\cite{bollobas2001random}, their extensions to the strong connectivity of random $k$-in(/out)-degree regular digraphs remain open. %We are unable to derive further useful results for our model from existing literature.
In the next section, we circumvent this difficulty by estimating the strong connectivity of random $k$-in-degree regular digraphs by that of random $p$-binomial digraphs (recall \cref{def:binD}) for appropriate $k$ and $p$.

\begin{remark}
	To draw connection between partitioning samplers and random $p$-binomial digraphs, consider the following ``binomial partitioning samplers'' construction modified from that of regular partitioning samplers:
	Instead of sampling a random $k$-subset of $C \setminus \set{s}$, the modified sampler includes each member of $C \setminus \set{s}$ into the ring independently with some fixed probability $p$.
	Correspondingly, a counterpart of~\cref{pro:Greg} for binomial partitioning sampler could be stated. For details, we refer to~\cref{app:binomial_samplers}.
\end{remark}

%% file: conjectures.tex
% !TEX root = main.tex

\section{Conjectures and Experiments}\label{sec:conjectures}

% We conjecture two inequalities about graphs being disconnected for different distributions, \ie for $\vec{\cG}^\regtag_{k,n}$ and $\vec{\cG}^\bintag_{p,n}$.
% These inequalities are chained together with results from previous sections to arrive at a closed-form upper bound for $\Pr{G \neq \core{G}}$, where $G$ is a random transaction graph induced by a (regular or binomial) partitioning sampler.
% Due to its nature, we can only provide empirical estimations on this upper bound.
% These estimations are close to the bound for different ring sizes and graph distributions, which further supports our conjectures.
%
% \todoI*{We give two conjectures on the distributions $\vec{\cG}^\regtag_{k,n}$ and $\vec{\cG}^\bintag_{p,n}$.
% Assuming that they hold, we chain them with the results from previous sections and arrive at a closed-form upper bound for $\Pr{G \neq \core{G}}$, where $G$ is a random transaction graph induced by a (regular or binomial) partitioning sampler.
% To support the conjectures, we provide numerical simulation results, which show that they indeed hold at least for all reasonable parameters in the context of cyptocurrencies.
% }

Towards finding the final piece of the puzzle of upper-bounding $\Pr{G \neq \core{G}}$ for $G$ induced by partitioning samplers, we put forth two conjectures concerning the probabilities of random $k$-in-degree regular digraphs and random $p$-binomial digraphs being strongly connected.
To gain confidence in these conjectures, we empirically estimate the probabilities for parameters which are reasonable in the context of cryptocurrencies.

\subsection{Conjectures}

Our first conjecture relates the two digraph distributions $\vec{\cG}^\regtag_{k,n}$ and $\vec{\cG}^\bintag_{k,n}$.

\begin{conjecture}\label{conj:reg_leq_bin}
	For $k, n \in \NN$ with $k < n$, and $p = \frac{k}{n-1} \leq 1$,
	\[
		\Pr[\vec{G} \gets \vec{\cG}^\regtag_{k,n}]{\vec{G} \notin \SC}
		\leq
		\Pr[\vec{G} \gets \vec{\cG}^\bintag_{p,n}]{\vec{G} \notin \SC}.
	\]
\end{conjecture}

Intuitively \cref{conj:reg_leq_bin} makes sense, since for all digraphs in the support of $\vec{\cG}^\regtag_{k,n}$, all nodes must be weakly connected to some other nodes (recall that all nodes have in-degree $k$), whereas this is not the case for $\vec{\cG}^\bintag_{p,n}$ with any $p<1$.

In search of a closed-form upper bound for $\Pr{G \neq \core{G}}$, we draw on the following result from Graham and Pike \cite{graham2008note}, which are developed based on the work of Pal{\'a}sti~\cite{palasti1966strong}.
% ``In search of'' is an idiom. Don't add ``the'' for the search.

\begin{lemma}[{\cite{graham2008note}}]\label{lem:limit}
	Let $c \in \RR$ be a constant and $p(n) := \frac{\ln n + c}{n}$.
	% Consider a $p$-binomial random digraph $\vec{G}$ of $N$ nodes, where %each of the $N^2$ possible edges (allowing self-loops) exists with probability
	% $p = p(N) = \frac{\ln N + c}{N}$, $c$ being an arbitrary, fixed real number. Let $\vec{\cG}$ be the distribution that $\vec{G}$ follows.
	It holds that
	\begin{equation*} \label{eq:limit}
		\lim_{n \to \infty} \Pr[\vec{G} \gets \vec{\cG}^{\bintag}_{p(n),n}]{\vec{G} \notin \SC} = 1 - e^{-2e^{-c}}.
	\end{equation*}
\end{lemma}

\begin{remark}
    Graham and Pike~\cite{graham2008note} considered a different model of digraphs where, unlike ours, self-loops are allowed. Their result however still holds under our model of digraphs, since self-loops have no effect on the strong connectivity of a digraph.
\end{remark}

% \begin{corollary} \label{cor:limit}
% 	Consider a transaction graph $\rvG$ induced by the binomial partitioning sampler, where $|R|$ rings are sampled among $|S|$ signers with decoy probability $p=\frac{\ln |R| + c}{|R|}$ for some $c \in \mathbb{R}$. Let $\cG$ be the distribution that $G$ follows, then
% 	\begin{align*}
% 		&\lim_{|R|\rightarrow \infty} \PrG{\rvG \neq \core{\rvG}} \\
% 		\leq& \lim_{|R|\rightarrow \infty} \PrG{\vec{\rvG}_u \notin \SC} \\
% 		=& 1 - e^{-2e^{-c}}.
% 	\end{align*}
% 	%	Consequently,
% 	%	\[
% 	%	\Pr{\core{G} \text{ is connected }} = \Omega\left(e^{-2e^{-c}}\right).
% 	%	\]
% \end{corollary}
%
% \begin{proof}
% 	The first inequality is by \cref{lem:boundGu} and \cref{lem:sc}, and the second equality by \cref{pro:Gbin} and \cref{lem:limit}.
% \end{proof}

\cref{lem:limit} moves us closer towards a closed-form upper bound for $\Pr{G \neq \core{G}}$, but unfortunately with two issues.
First, the results of Pal{\'a}sti~\cite{palasti1966strong} and Graham and Pike~\cite{graham2008note} seem to crucially rely on setting $p(n) := \frac{\ln n + c}{n}$,
and infer nothing about the case with general $p$.
Second, their results concern only about infinite digraphs, but say little about finite digraphs.

To close the gaps, we propose our second conjecture, which is obtained heuristically by
plugging in $c = pn - \ln n$ back to the limit in \cref{lem:limit}.

\begin{conjecture} \label{conj:bin_upper_bound}
	For any probability $p \in [0,1], n \in \NN$,
	\[
		\Pr[\vec{G} \gets \vec{\cG}^\bintag_{p,n}]{\vec{G} \notin \SC}
		\leq
		1 - e^{-2e^{\ln n - pn}}.
	\]
\end{conjecture}

While we are unable to provide analytical proofs, both of the conjectures hold for all choices of $(k,n)$ in our numerical simulations in~\cref{sec:exp}, where $(k,n)$ are chosen to be realistic in the context of cryptocurrencies.

Finally, taking these two conjectures, we can bridge the established results and arrive at the concluding statement below.

\begin{corollary}\label{cor:final}
    Let $U$ be a set of users and $P$ be a partition of $U$.
    Let $k \in \NN$ such that $k < |C|$ for each $C \in P$.
	Let $n := \max_{C \in P} |C|$.
	If~\cref{conj:reg_leq_bin,conj:bin_upper_bound} hold, then for any $m \leq |U|$,
	\[
		\Pr[G \gets \cG^{\RegSamp[P, k]}(U,1^m)]{G \neq \core{G}} \leq |P| \tuple{1 - e^{-2e^{\ln n - k}}}.
	\]
\end{corollary}

\begin{proof}
	\qquad\quad $\begin{aligned}[t]
	\Pr[G \gets \cG^{\RegSamp[P, k]}(U,1^m)]{G \neq \core{G}} 
	\leq& \sum_{C \in P} \Pr[\vec{G} \gets \vec{\cG}^\regtag_{k,|C|}]{\vec{G} \notin \SC} \\
	\leq& \sum_{C \in P} \Pr[\vec{G} \gets \vec{\cG}^\bintag_{p(C),|C|}]{\vec{G} \notin \SC} \\
	\leq& \sum_{C \in P} \tuple{1 - e^{-2e^{\ln |C| - \frac{k}{|C|-1}|C|}}} \\
	<& \sum_{C \in P} \tuple{1 - e^{-2e^{\ln |C| - k}}} \\
	\leq& |P| \tuple{1 - e^{-2e^{\ln n - k}}},
	\end{aligned}$\\
	where the first inequality follows from \cref{cor:Greg},
	the second follows from \cref{conj:reg_leq_bin} by setting $p(C) = \frac{k}{|C|-1}$ for $C \in P$,
	and the third follows from \cref{conj:bin_upper_bound}.
\end{proof}

\subsection{Experiments}
\label{sec:exp}

To support our conjectures, we empirically estimated the probabilities
\begin{align*}
    p^{\regtag}_{k,n} &:= \Pr[\vec{G} \gets \vec{\cG}^{\regtag}_{k,n}]{\vec{G} \notin \SC}~\text{and} \\
    p^{\bintag}_{k,n} &:= \Pr[\vec{G} \gets \vec{\cG}^{\bintag}_{p,n}]{\vec{G} \notin \SC},
\end{align*}
where $p = \frac{k}{n-1}$, for values of $k$ ranging from 1 to 16 and values of $n$ from $2^2$ to $2^{12}$ in exponential steps.
In each case we sampled $8000$ graphs, verified whether $\vec{G} \notin \SC$, and compared the average with the upper bound
$$\bar{p}^{\bintag}_{k,n} := 1-e^{-2e^{\ln n - \frac{k}{n-1}n}}$$
in~\cref{conj:bin_upper_bound} when setting $p = \frac{k}{n-1}$.

\begin{figure*}[t]
	\centering
	\includegraphics[width=.47\textwidth]{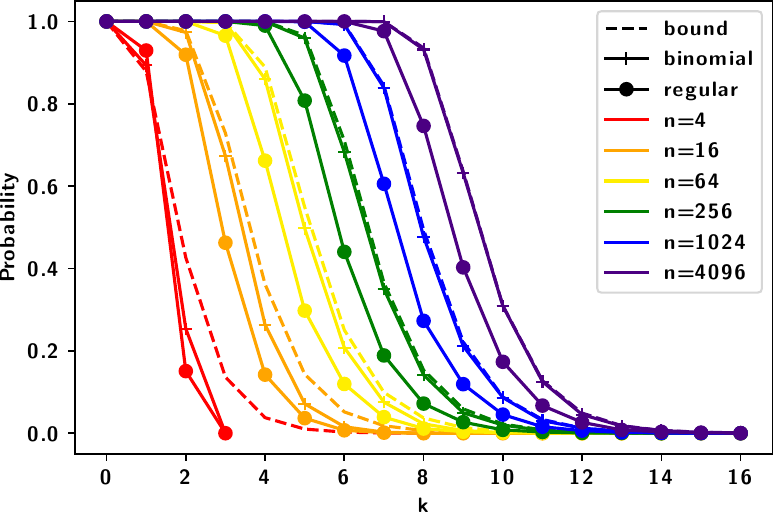}\hspace{.5cm}
	\includegraphics[width=.47\textwidth]{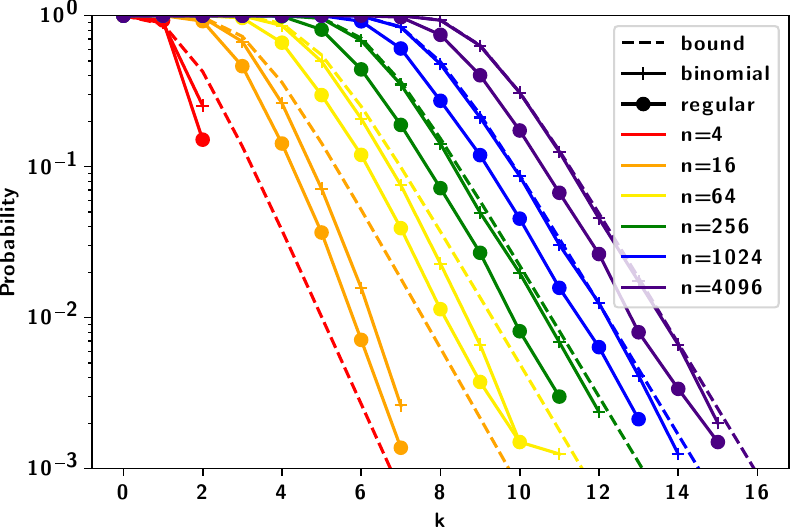}
	\caption{Plots of $p^{\regtag}_{k,n}$, $p^{\bintag}_{k,n}$, and $\bar{p}^{\bintag}_{k,n}$ against $k$ for selected values of $n$ in both linear- and log-scale.}
	\label{fig:plot:universes}
\end{figure*}

\begin{figure*}[t]
  \centering
  \includegraphics[width=.47\textwidth]{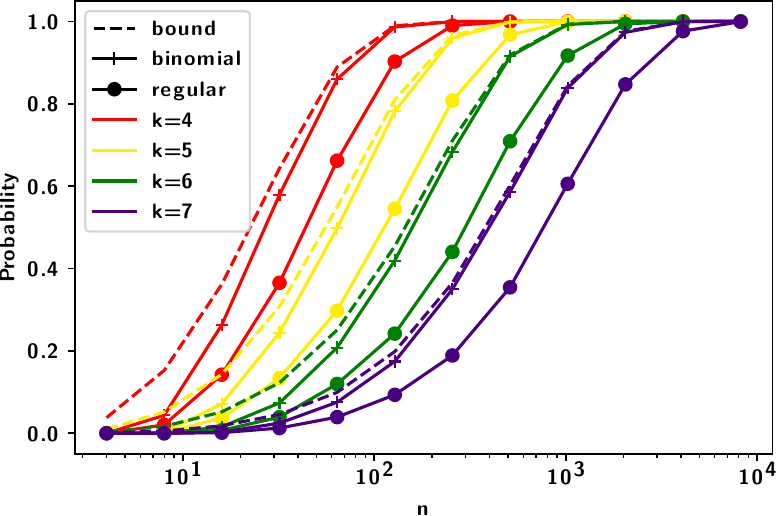}\hspace{.5cm}
  \includegraphics[width=.47\textwidth]{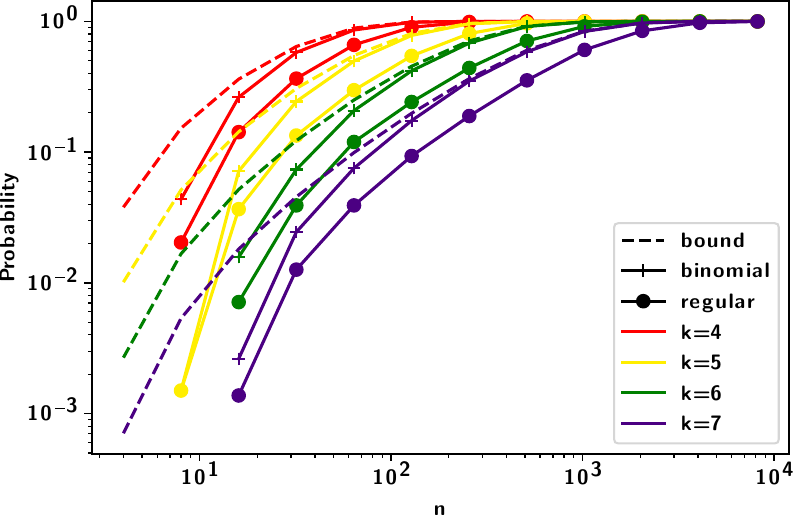}
  \caption{Plots of $p^{\regtag}_{k,n}$, $p^{\bintag}_{k,n}$, and $\bar{p}^{\bintag}_{k,n}$ against $n$ for selected values of $k$ in both linear- and log-scale.}
  \label{fig:plot:rings}
\end{figure*}

In~\cref{fig:plot:universes}, we plotted $p^{\regtag}_{k,n}$ (dot mark, ``regular''), $p^{\bintag}_{k,n}$ (plus mark, ``binomial''), and $\bar{p}^{\bintag}_{k,n}$ (dashed, ``bound'') against $k$ for different values of $n$ in both linear- and log-scale.
In the log-scale plot, values smaller than $10^{-3}$ are omitted for their instability due to the limited sampling size.
Similarly, in~\cref{fig:plot:rings} we plotted the same values against $n$ for different values of $k$.
%
%
% The probability of $\vec{G} \notin \SC$ against $k$ and $n$ are plotted in \cref{fig:plot:universes,fig:plot:rings} respectively.
% In~\cref{fig:plot:universes} we additionally plotted the same curves but in log-scale, where
% In both figures, the upper bounds in~\cref{conj:bin_upper_bound} are plotted as dashed lines and are labeled ``limit''.
% The curves for $\vec{\cG}^{\regtag}_{k,n}$ and $\vec{\cG}^{\regtag}_{p,n}$ we show the limit by dashed lines and distinguish the samplers by the plot markers.
% We color code the choices of $n$ (resp. $k$).

From \cref{fig:plot:universes,fig:plot:rings}, we observe that both conjectured upper bounds appear to hold in general. Upon closer inspection, on the one hand, we observe that the first bound
\[p^{\regtag}_{k,n} \leq p^{\bintag}_{k,n}\]
 becomes tighter as the number of nodes $n$ decreases. This makes sense since the variance of the in-degree of the nodes in the graphs sampled from $\vec{\cG}^{\bintag}_{p,n}$ decreases as $n$ decreases. On the other hand, we notice that the second conjectured upper bound
\[p^{\bintag}_{k,n} \leq \bar{p}^{\bintag}_{k,n}\]
 becomes tighter as $n$ increases. This is also expected as the bound was heuristically derived from the limit of $p^{\bintag}_{k,n}$ as $n$ tends to infinity.

% the probabilities for the binomial sampler closely resembles the limit for large universe sizes, while for small universes the conjectured bound becomes less tight.
% For the regular samplers, the empirical values suggest that reducing $k$ by one compared to the limit will still give comparable results.
% \Cref{fig:plot:rings} confirms the intuition that, for large $n$ the success probability of an attack increases but remains within the conjectured bound.

%% file: recommendation.tex
% !TEX root = main.tex

\section{Interpretation of Our Results}
\label{sec:interpretation}

We conclude our work by stating a ring size for partitioning samplers which is sufficient to defeat graph analysis. We also discuss how our results extend to the setting with an active adversary, who attempts to deanonymise honest signers by injecting fake ones in the so-called ``black marble attacks''~\cite{MRL-0001,MRL-0004,8456034}.

\subsection{On Defeating Graph Analysis}

We discuss what our results mean in the context of (passive) graph-based deanonymisation attacks.
We begin by showing that, for transaction graphs $G$ induced by $k$-regular partitioning samplers, conditioned on $G = \core{G}$, the trivial deanonymisation strategy described in~\cref{sec:ring_samp_defs} is the best strategy.

\begin{lemma}\label{lem:advantage_bound_regular}
	Let $k \in \NN$, $U$ be a set of users, $P$ be a partition of $U$ where $|C|> k$ for each $C \in P$.
	Let $\Samp = \RegSamp[P,k]$.
	For any adversary $\adv$,
	\begin{equation*}
		\Pr{\mathsf{Exp}_{\adv, \Samp}(U)}
		\leq \Pr{G \neq \core{G}} + \frac{1}{k+1}
	\end{equation*}
	where the probabilities are taken over the randomness of $\adv$ and $(G,M) \gets \cG^{\Samp}(U,1^{|U|})$.
\end{lemma}

\begin{proof}
	Observe that
	\begin{align*}
		&\Pr{\mathsf{Exp}_{\adv, \Samp}(U)} \\
		\leq &\Pr{G \neq \core{G}} + \Pr{\mathsf{Exp}_{\adv, \Samp}(U) \middle | G = \core{G}},
	\end{align*}
%	\textcolor{OliveGreen}{
	which is obtained by applying the law of total probability and upper-bounding two probability terms by 1. %}
	Then, it suffices to show that
	\[
		\Pr{\mathsf{Exp}_{\adv, \Samp}(U) \middle | G = \core{G}} \leq \frac{1}{k+1}.
	\]

	\noindent Consider the distribution
	\[
	\hat{\cG}^\Samp := \set{
		(G,M):
		\begin{aligned}
			&(G,M) \gets \cG^\Samp(U,1^{|U|}) \\
			&G = \core{G}
		\end{aligned}
	}.
	\]
	% {\color{OliveGreen}
    In other words, $\hat{\cG}^\Samp$ is the same as $\cG^\Samp$ conditioning on $G = \core{G}$.
	Let $\hat{\mathsf{Exp}}_{\adv, \Samp}$ be the same as $\mathsf{Exp}_{\adv, \Samp}$, except that the procedure $({G}, {M}) \gets {\cG}^\Samp$ is replaced by $(\hat{G}, \hat{M}) \gets \hat{\cG}^\Samp$.
	We can rewrite
	\[\Pr{\mathsf{Exp}_{\adv, \Samp}(U) \middle | G = \core{G}} = \Pr{\hat{\mathsf{Exp}}_{\adv, \Samp}(U)}.\]
	Since $\Samp$ is a partitioning sampler, for any fixed members of any fixed ring, the probability of the ring being sampled by each member is the same. That is, %for any pair of signers $s$ and $s'$, %edges $(s,r)$ and $(s',r)$ with the same ring $r$ in $\hat{G}$, 	
	for any fixed ring $r$ in the support of $\bigcup_{u \in U} \Samp(U,u)$, and any fixed $s, s' \in r$, it holds that
	\[\Pr{r = \Samp(s)} = \Pr{r = \Samp(s')}.\]
	Therefore, conditioned on the event $E = ((s,r),(s',r) \in \hat{G})$ for any fixed $r,s,s'$,
	\[\Pr{(s,r) \in \hat{M} | E} = \Pr{(s',r) \in \hat{M} | E}\] 
	with probabilities taken over $(\hat{G}, \hat{M}) \gets \hat{\cG}^\Samp$. %}
	Consequently, for any edge $(s^*,r^*)$ output by $\adv(\hat{G})$,
	% {\color{OliveGreen}
    \begin{align*}
	\Pr{\hat{\mathsf{Exp}}_{\adv, \Samp}(U)} 
	=~&\Pr{(s^*,r^*) \in \hat{M}} \\ 
	=~&\Pr{(s^*,r^*) \in \hat{G}} \Pr{(s^*,r^*) \in \hat{M} | (s^*,r^*) \in \hat{G}} \\
	&\quad + \Pr{(s^*,r^*) \notin \hat{G}} \Pr{(s^*,r^*) \in \hat{M} | (s^*,r^*) \notin \hat{G}}  \\
	=~&\Pr{(s^*,r^*) \in \hat{G}} \cdot \frac{1}{|r^*|} + 0 \leq \frac{1}{k+1}, 
	\end{align*}
	as desired.%}
\end{proof}

From~\cref{lem:advantage_bound_regular}, if the parameters $P$ and $k$ are set such that $\Pr[G \gets \cG^{\RegSamp[P,k]}(U,1^{|U|})]{G \neq \core{G}} \leq \frac{1}{k+1}$, then $\RegSamp[P,k]$ is $\epsilon$-secure against graph-based deanonymisation attacks for $\epsilon = \frac{2}{k+1} = O(1/k)$, which is optimal up to a constant factor of $2$.
In the next theorem, we give a sufficient condition on $k$ with which this holds.

\begin{theorem}\label{thm:concrete_security}
	Let $k, n \in \NN$, $U$ be a set of users, and $P$ be a partition of $U$ where $|C| = n > k$ for each $C \in P$.
	If~\cref{conj:reg_leq_bin,conj:bin_upper_bound} hold and
	\[k \geq \ln (2|U|) + \sqrt{2\ln (2|U|)},\]
	then $\RegSamp[P,k]$ is $\frac{2}{k+1}$-secure against graph-based deanonymisation attacks.
\end{theorem}

\begin{proof}
	By~\cref{lem:advantage_bound_regular}, it suffices to show that $\Pr[G \gets \cG^{\RegSamp[P,k]}(U,1^{|U|})]{G \neq \core{G}} \leq \frac{1}{k+1}$ for the given parameters.
	Let $k':= k+1$.
	If~\cref{conj:reg_leq_bin,conj:bin_upper_bound} hold, then by \cref{cor:final} it suffices to set up parameters such that
	\[
		|P| \tuple{1 - e^{-2e^{\ln n - k}}} \leq \frac{1}{k'}
	\]
	or equivalently
	\[
		k \geq \ln \left(\frac{-2n}{\ln\left(1-\frac{1}{|P|k'}\right)}\right),
	\]
	where $n = \max_{C\in P} |C|$ is the maximum chunk size. %Applying the inequality $\ln(1+x)\leq x$ for $x>-1$ to
	Since $\ln \left(1-\frac{1}{|P|k'}\right) \leq -|P|k'$, we have
	\[
	k \geq \ln \left(2n|P|k'\right)
	\]
	as a sufficient condition, so it suffices to solve $k'$ for
	\begin{align*}
	k' - \ln k' \geq \ln (2en|P|),
	\end{align*}
	the solution of which is
	\begin{align*}
	k' \geq -W_{-1}\left(\frac{-1}{2en|P|}\right),
	\end{align*}
	where $W_{-1}(\cdot)$ is the Lambert W function of branch $-1$. From \cite{journals/icl/Chatzigeorgiou13} we know that
	\[
	-1-\sqrt{2x}-x < W_{-1}\left(-e^{-x-1}\right)
	\]
	for all $x>0$. Substituting $x=\ln (2n|P|)$, we conclude that it suffices to set
	\[k \geq \ln (2n|P|) + \sqrt{2\ln (2n|P|)}.\]
	In the particular case stated in the theorem statement, where the set of users is partitioned into chunks of equal size $n$, \ie $|U| = n|P|$, it suffices to set
	\begin{equation*}
		k \geq \ln (2|U|) + \sqrt{2\ln (2|U|)}. \qedhere
	\end{equation*}
\end{proof}

  For concreteness, suppose it is believed that the number of all users $|U|$ will never exceed $2^{64}$, then~\cref{thm:concrete_security} suggests that, by setting the number of dummies $k$ to at least $55$, the probability that an adversary identifies a signer is at most $\frac{2}{k+1} \leq \frac{1}{28}$.
  Suppose that users are comfortable with a 1-in-$t$ anonymity for some $t \geq 28$, then it should suffice to set $k$ as such that $\frac{k+1}{2} = t$, yielding a ring size of $2t$.

  %\textcolor{OliveGreen}{
	In the example of Monero, its current recommended ring size of 11 seems far too small under our model. We note, however, that a ``correct'' level of anonymity is itself a subjective matter. If a Monero user is willing to accept that the anonymity set size will be reduced, say, from 11 to 6 and is comfortable with an anonymity set of size 6, then 11 might still be an acceptable choice. Future empirical study on the actual reduction in anonymity of Monero users could offer useful insights in this direction.%}

  We remark that the above recommendation for the ring size is conservative for several reasons.
  First, the upper bound of the adversary's success probability given in~\cref{lem:advantage_bound_regular} is loose in the sense that, while we let $\Pr{\mathsf{Exp}_{\adv, \Samp}(U) \middle | G \neq \core{G}} \leq 1$ in its derivation, having $G \neq \core{G}$ does not necessarily mean that the adversary immediately has a drastic advantage.
  Rather, we believe that the anonymity degrades gracefully depending on how close $\core{G}$ is to $G$.
  Second, ring samplers which are secure in our model resist even untargeted attacks against individual signatures.
  In practice, being able to identify the signer of one random signature does not seem very useful, especially in the LRS setting where each signing key is only used once.
  A more meaningful attack, say in the setting of cryptocurrencies, is to identify the signers of a chain of $\ell > 1$ transactions.
  However, the probability of successfully doing so intuitively decreases exponentially in $\ell$.

\subsection{On Black Marble Attacks}

A type of active deanonymisation attacks is the so-called ``black marble attacks''~\cite{MRL-0001,MRL-0004,8456034}, where the adversary actively injects signers, called black marbles, into the set of users, such that including them in rings do not contribute towards the anonymity of honest signers. In the context of cryptocurrencies, injecting black marbles often incur a monetary cost. It is therefore reasonable to assume that the adversary is only able to inject a bounded number of black marbles per some unit of time. Such attacks can be captured by the experiment $\mathsf{Exp}_{\adv,\Samp,\pi}$ in~\cref{def:graph-based-anon}.
% \begin{enumerate}
% 	\item The adversary specifies a subset $A \subseteq U$ of black marbles.
% 	\item Sample a transaction graph $G = (U,U,E) \gets \cG^\Samp$ induced by a ring sampler $\Samp$.
% 	\item Remove nodes and edges corresponding to the black marbles from $G$, \ie compute $G^- = (U^-, U^-, E^-)$ where $U^- := U \setminus A$ and $E' := E \cap (U^- \times U^-)$.
% 	\item Sample random permutations $\pi_0, \pi_1: U^- \to U^-$.% where $U^* := [|U^-|]$.
% 	\item Compute the bipartite graph $G^* = (U^-,U^-,E^*)$ where $E^* := \set{(\pi_0(i), \pi_1(j)): (i,j) \in E^-}$.
% 	\item Run the adversary on $G^*$ and obtain as output a tuple $(i^*, j^*)$.
% 	\item Output ``win'' if $(i^*, j^*) = (\pi_0(j), \pi_1(j))$ for some $j \in U^-$.
% \end{enumerate}
%As before, we clearly have $G = \core{G}$ if and only if $G^* = \core{G^*}$.

For simplicity, suppose that each chunk $C \in P$ is of size $|C| = n$ and contains $\beta \cdot n$ black marbles for some $\beta \in [0,1]$. Then the ``effective'' number of users (in the sense of providing anonymity) is given by $(1-\beta) \cdot |U|$. This is captured by a predicate $\pi$ which checks that $|B \cap C| \leq \beta \cdot |C|$ for all $C \in P$.

Suppose that $\Samp = \RegSamp[P,k]$. Notice that, after removing the black marbles, the induced digraphs of the chunks of the transaction graphs $G$ are no longer $k$-in-degree regular, and are tedious to analyse.
Fortunately, for the case with binomial partitioning sampler (detailed in \cref{app:binomial_samplers}), we observe that the induced digraphs of the chunks of the transaction graphs $G$ follow the distributions $\set{\vec{\cG}^{\bintag}_{p,(1-\beta) \cdot n}}_{C \in P}$. That is, injecting black marbles only decreases the size parameter of the $p$-binomial digraph distribution by a factor of $(1-\beta)$. We can therefore still apply~\cref{conj:bin_upper_bound} and obtain an analogous upper bound for this setting.
By replacing $|U|$ with $(1-\beta) \cdot |U|$ in the proof of~\cref{thm:concrete_security}, we conclude that it suffices to set
\[
	p \geq \frac{\ln (2 \cdot (1-\beta) \cdot |U|) + \sqrt{2\ln(2 \cdot (1-\beta) \cdot |U|)}}{(1-\beta)n - 1}
\]
to defeat graph analysis.
Revisiting the setting of $\Samp = \RegSamp[P,k]$, the above heuristically suggests that
\[
	k \gtrsim \frac{\ln (2 \cdot (1-\beta) \cdot |U|) + \sqrt{2\ln(2 \cdot (1-\beta) \cdot |U|)}}{1-\beta}
\]
suffices to defeat graph analysis.

%% file: app-entropy.tex
% !TEX root = main.tex

\section{Entropy-based Anonymity}\label{app:entropy}

Ronge~\etal~\cite{PoPETs:RELSY21} introduced an anonymity measure for ring samplers based on conditional min-entropy.
They also proved that the regular partitioning sampler achieves close to optimal anonymity with respect to this measure under a realistic assumption about the signer distribution.
Here we recall the definition of this measure, and prove that the binomial partitioning sampler also achieves close to optimal anonymity with respect to this measure under the same assumption.
In this context, a ring sampler $\Samp$ is assumed to always sample a ring for some signer $s$ from the set of users, for brevity we omit in the following the input $U$ and write simply $\Samp(s)$.

%We recall the definitions of conditional min-entropy, signer distributions, and anonymity of ring samplers (in the sense of~\cite{PoPETs:RELSY21}).

% Min-entropy is the most pessimistic measure for the unpredictability of the outcome. This can be used to define conditional min-entropy as following.

\begin{definition}[Conditional Min-entropy]
	Let $\cX$ and $\cY$ be discrete distributions with probability mass functions $p_\cX$ and $p_\cY$ respectively.
	Let $p_{\cX|\cY}$ and $p_{\cY|\cX}$ be the corresponding conditional probability mass functions.
	The conditional min-entropy of $\cX$ given $\cY$ is defined as
	\begin{align*}
		H_{\infty}(\cX |\cY)
		:= &- \ln \left( \sum_y  p_\cY(y) \cdot \max_x p_{\cX|\cY}(x|y) \right) \\
		= &- \ln \left( \sum_y \max_x \tuple{p_{\cY|\cX}(y|x) \cdot p_\cX(x)}  \right).
	\end{align*}
\end{definition}

% To give meaning to \cref{def:ringsampler} and \cref{def:anon}, it is necessary to define the term \enquote{signer distribution}.

\begin{definition}[Signer Distributions~\cite{PoPETs:RELSY21}]\label{app:sigdist}
	A signer distribution $\sdv$ is a distribution over $2^U\setminus\set{\emptyset}$, \ie each sample of $\sdv$ is a non-empty subset of $U$.
	If all samples of $\sdv$ are singletons, \ie $\Pr[S \gets \sdv]{|S| = 1} = 1$, we say that $\sdv$ is a single-signer distribution.
\end{definition}

% This can be used to define an anonymity measure.

\begin{definition}[Anonymity~\cite{PoPETs:RELSY21}]\label{def:anon}
	The anonymity of $\Samp$ with respect to a signer distribution $\sdv$ is defined as
	\[
	\alpha[\sdv, \Samp] := H_\infty(\sdv|\Samp(U,\sdv)).
	\]
\end{definition}

%Intuitively, anonymity of a ring sampler should be a measure of \enquote{how good} the sampler is, \ie provide a number that makes samplers comparable.
%Further, at least as long a sampler does not depend on the actual signer distribution (see \cref{app:sigdist}) the measure needs to take it into account as otherwise the quality could differ a lot for different distributions and therefore would be not useful.

%To address this, the measure in \cite{PoPETs:RELSY21} uses min-entropy\footnote{which is the negative logarithm of the probability an adversary guesses the signer(s) correctly} as a most pessimistic base for the adversaries winning chance conditioned on the observed ring.
%This measure is upper bounded by the min-entropy itself.

Note that the anonymity measure defined in~\cref{def:anon} captures only ``local'' anonymity since it disregards information about the signer leaked from the rings generated by other users.
While the anonymity measure could be generalised to the ``global'' setting by simply considering the min-entropy of $\sdv$ conditioned on a sequence of rings, analysing ring samplers with respect to such generalised measure appears to be difficult.
Indeed, all analyses done in~\cite{PoPETs:RELSY21} were with respect to the local measure defined in~\cref{def:anon}.

Ronge~\etal~\cite{PoPETs:RELSY21} proved that the regular partitioning samplers achieve close to optimal anonymity with respect to the above measure under a mild assumption.

% For the regular ring sampler $\RegSamp$, a lower bound on the anonymity can be given.

\begin{lemma}[{\cite[Theorem~6.3]{PoPETs:RELSY21}}]
	Let $P$ be a partition of $U$.
	Let $\sdv$ be a single-signer distribution with probability mass function $p_\sdv$.
	For each $C \in P$, let $\mu_C$ be the mean of $p_\sdv(s)$ over all $s \in C$, \ie $\mu_C := |C|^{-1} \sum_{s \in C} p_\sdv(s)$.
	Suppose that for all $C \in P$, all $s \in C$, it holds that $|p_\sdv(s) - \mu_C| \leq \epsilon_C$ for some $\epsilon_C \geq 0$.
	Let $\epsilon_P := \sum_{C \in P} |C| \epsilon_C$. Then
	\[
	\alpha(\sdv, \RegSamp[P, k]) > \ln k - \ln (\epsilon_P + 1).
	\]
\end{lemma}

\section{Binomial Partitioning Samplers}\label{app:binomial_samplers}

%Being aware of the hardness of analysing regular partitioning samplers, we define a new type of partitioning sampler -- the binomial partitioning samplers --
%in the hope that analysing it leads to a candidate upper bound for $\Pr{G\neq\core{G}}$ for regular partitioning samplers.

Similar to \cref{pro:Greg} which relates the regular partitioning samplers to the distribution $\vec{\cG}^\regtag_{k,n}$, we can construct a new type of partitioning samplers -- the binomial partitioning samplers -- which could be related to the distribution $\vec{\cG}^\bintag_{p,n}$.

Loosely speaking, a binomial partitioning sampler similarly partitions the set of users into chunks, and within each chunk the sampler includes each signer as decoy in a ring with some fixed probability independent of all other signers. The independence of signers being chosen as decoys turns out to make the analysis of the corresponding induced transaction graphs much easier than that of the regular partitioning samplers.

As in \cref{sec:sampreg}, we consider the case where there is only one public partition of $U$ and only one signer per ring. A binomial partitioning sampler $\BinSamp[P,p]$, parametrised by the partition $P$ of $U$ and a decoy probability $p$, is defined as follows.

\begin{quote}
	$\BinSamp[P,p](U,s)$:
	Initiate $r := \set{s}$.
	Let $C \in P$ be the unique chunk containing $s$ and, for each $d \in C \setminus \set{s}$, run $r := r \cup \set{d}$ with probability $p$.
	Output $r$.
\end{quote}

In the setting where $s$ is a set of signers instead of a single one, a ring could be sampled by repeating the above procedure for each member of $s$ and taking the union.

In case
$\sdv$ is a single-signer distribution,
$|C| = n$ for each chunk $C \in P$, and $p=\frac{k}{n-1}$, the binomial partitioning samplers $\BinSamp[P,p]$ are analogous to the regular partitioning samplers $\RegSamp[P,k]$, in the sense that the former has expected ring size $k+1$ while the latter has fixed size $k+1$. Furthermore, the numbers of decoys in a ring sampled from $\BinSamp[P,p]$ follow the binomial distribution with mean $k$ and variance $k(1-p)$.

Similar to the regular partitioning samplers, the distribution of transaction graphs induced by a binomial partitioning sampler is related to some specific distribution. Clearly, the distribution $\cG^{\BinSamp[P,p]}$ can be partitioned as $\set{\cG_C}_{C \in P}$, each $\cG_C$ being independent of each other and representing the distribution of induced transaction graphs of a chunk $C\in P$. Furthermore, each $\cG_C$ can be sampled by setting each of the possible edges independently with probability $p$. We therefore arrive at the following analogy to \cref{pro:Greg}.

% Now we observe that, since within each chunk each signer is being chosen as decoy in a ring with probability $p$ independent of all other signers, the distribution of $\ID{\cG^{\BinSamp[P,p]}_{C,C}}$ is the same as $\vec{\cG}^\bintag_{p,|C|}$ for each $C\in P$.

% \begin{proposition} \label{pro:Gbin}
	% %	Let $P$ be such that $|C| = n$ for each $C \in P$.
	% 	Consider the partition $\set{\cG^{\BinSamp[P,p]}_{C,C}}_{C \in P}$ of $\cG^{\BinSamp[P, p]}$.
	% 	It holds that $\ID{\cG^{\BinSamp[P,p]}_{C,C}} = \vec{\cG}^\bintag_{p,|C|}$ for each $C \in P$.
	% \end{proposition}

%In the above, $\ID{\cG^{\BinSamp[P,p]}_{C,C}} = \vec{\cG}^\bintag_{p,|C|}$ means that sampling $\vec{G}$ from $\vec{\cG}^\bintag_{p,|C|}$ is equivalent to first sampling $G$ from $\cG^{\BinSamp[P,p]}_{C,C}$ and then computing $\vec{G} = \ID{G}$.

\begin{lemma}\label{pro:Gbin}\label{cor:Gbin}
	Let $U$ be a set of users and $P$ be a partition of $U$.
	Let $p \in [0,1]$.
	Write $\Samp := \BinSamp[P, p]$.
	For any $m \leq |U|$, %it holds that
	\[
	\Pr[G \gets \cG^{\Samp}(U,1^m)]{G \neq \core{G}}
	\leq \sum_{C \in P} \Pr[\vec{G} \gets \vec{\cG}^\bintag_{p,|C|}]{\vec{G} \notin \SC}.
	\]
\end{lemma}

\begin{proof}
	Similar to the proof of \cref{cor:Greg}.
\end{proof}

Analogous to~\cref{lem:advantage_bound_regular}, it is not difficult to show a similar bound for $\Samp = \BinSamp[P,p]$. As the binomial partitioning sampler has variable ring sizes, in the analysis we need to use a tail bound to argue that, with overwhelming (in $k$) probability, all rings produced by $\cG^{\BinSamp[P,k]}$ have size not far from $k+1$. Since the argument is tedious but straightforward, we omit it.

For the sake of completeness, we analyse the anonymity of the binomial partitioning samplers according to the entropy-based measure. It turns out that the binomial partitioning samplers have the same near-optimal level of anonymity as the regular partitioning samplers do.

\begin{theorem} \label{thm:entropy}
	Let $P$ be a partition of $U$.
	Let $\sdv$ be a single-signer distribution with probability mass function $p_\sdv$.
	Let $P$ and $k \in \NN$ be such that $p|C| > k$ for each $C \in P$.
	For each $C \in P$, let $\mu_C$ be the mean of $p_\sdv(s)$ over all $s \in C$, \ie $\mu_C := |C|^{-1} \sum_{s \in C} p_\sdv(s)$. Suppose that for all $C \in P$, all $s \in C$, it holds that $|p_\sdv(s) - \mu_C| \leq \epsilon_C$ for some $\epsilon_C \geq 0$. Let $\epsilon_P := \sum_{C \in P} |C| \epsilon_C$. Then
	\[
		\alpha(\sdv, \BinSamp[P, p]) > \ln k - \ln (\epsilon_P + 1).
	\]
\end{theorem}

\begin{proof}
	Let $\Samp = \BinSamp[P, p]$.
	For any $s \in U$, as the chunk containing $s$ is unique, we know that $\bigcup_{C \in P} (2^C \setminus \set{\emptyset})$ is a superset of the collection of all possible rings.
	Write $\cR_C := 2^C \setminus \set{\emptyset}$ and $\cR := \bigcup_{C \in P} \cR_C$.
	Since the ring given by the sampler must contain the signer, we have for all signer $s$ and for all $r \in \cR$,
	\begin{equation*}
		\Pr{\Samp(U,s) = r \land s \notin r} = 0.
	\end{equation*}
	If $s \in C \in P$, then each element in $C \setminus \set{s}$ has a probability $p$ to be included in $r \setminus \set{s}$.
	On the other hand, if $s \notin C \in P$, then we must have $r \notin \cR_C$.
	Therefore, for any $s \in U$, $C \in P$, and $r \in \cR_C$, we have
	\begin{align*}
		&\Pr{\Samp(U,s) = r \land s \in r} \\
		=&
		\begin{cases}
			p^{|r|-1} (1-p)^{(|C|-1) - (|r|-1)} & s \in C\\
			0 & s \notin C
		\end{cases} \\
		=&
		\begin{cases}
			p^{|r|-1} (1-p)^{|C|-|r|} & s \in C\\
			0 & s \notin C.
		\end{cases}
	\end{align*}
	Now, we analyse the anonymity of the sampler.
	\begin{align*}
		&2^{-\alpha[\sdv,\Samp]}
		= 2^{H_\infty(\sdv|\Samp(U,\sdv))} \\
		= &\sum_{r \in \cR}
			\max_{s \in U} \tuple{p_{\Samp(U,\sdv)|\sdv}(r|s) \cdot p_\sdv(s)} \\
		\leq &\sum_{C \in P} \sum_{r \in \cR_C}
			\max_{s \in C} \tuple{\Pr{\Samp(U,s) = r \land s \in r} \cdot p_\sdv(s)} \\
		= &\sum_{C \in P} \sum_{r \in \cR_C} p^{|r|-1}
			(1-p)^{|C|-|r|} \max_{s \in C} p_\sdv(s) \\
		= &\sum_{C \in P} \frac{1 - (1-p)^{|C|}}{p}
			\max_{s \in C} p_\sdv(s) \\
		\leq &\sum_{C \in P} \frac{1 - (1-p)^{|C|}}{p} (\mu_C + \epsilon_C) \\
		< &\sum_{C \in P} \frac{|C|}{k} (\mu_C + \epsilon_C) \\
		= &\frac{\epsilon_P + 1}{k}. \qedhere
	\end{align*}
\end{proof}